\documentclass[]{aa}

\usepackage{graphicx}             
\usepackage{float}                 
\usepackage{placeins}              
\usepackage{adjustbox}             
\usepackage{lscape}                
\usepackage{wrapfig}               
\usepackage{epstopdf}              
\usepackage{subfigure}             
\usepackage{booktabs}              
\usepackage{amsmath}               
\usepackage[T1]{fontenc}           
\usepackage{float}                 
\usepackage{gensymb}               
\usepackage{rotating}              
\usepackage{siunitx}
\usepackage{dcolumn}
\usepackage{xcolor}

\usepackage{txfonts}
\usepackage{natbib}

\bibpunct{(}{)}{;}{a}{}{,} 

\DeclareRobustCommand{\VAN}[3]{#2}
\let\VANthebibliography\thebibliography
\def\thebibliography{\DeclareRobustCommand{\VAN}[3]{##3}\VANthebibliography}

\DeclareGraphicsExtensions{.ps}
\usepackage{graphicx}
\usepackage{txfonts}
\begin{document}

\title{Revealing the state transition of Cen X-3 at high spectral resolution with \textit{Chandra}. }

\author{
Sanjurjo-Ferr\'{i}n, G.$^{1}$,
Torrej\'on, J.M.$^{1}$,
Oskinova, L.$^{2}$,
Postnov, K.$^{3}$,
Rodes-Roca, J.J.$^{1}$,
Schulz, N.$^{4}$,
Nowak, M.$^{5}$
}

\institute{
$^{1}$Instituto Universitario de F\'{i}sica Aplicada a las Ciencias y las Tecnolog\'ias, Universidad de Alicante, 03690 Alicante, Spain\\
$^{2}$Institute for Physics and Astronomy, Universit\"{a}t Potsdam, 14476 Potsdam, Germany\\
$^{3}$ Sternberg Astronomical Institute, Moscow M.V. Lomonosov State University, Universitetskij pr, 13, Moscow 119234, Russia\\
$^{4}$ MIT Kavli Institute for Astrophysics and Space Research, Cambridge, Massachussetts, USA\\
$^{5}$ Department of Physics, Washington University in St. Louis, Missouri, USA
}

\date{Received XXX / Accepted XXX}

\abstract{Cen X-3 is a compact, high-mass X-ray binary (HMXRB), likely powered by Roche lobe overflow. We present a phase-resolved X-ray spectral and timing analysis of a target of opportunity \textit{Chandra} observation made during a low-flux to high-flux transition. The high-resolution spectra allow us to delve into the events that occurred during this episode. The spectrum is described by a single black body absorbed by a local column density of the order of $10^{23-24}$ cm$^{-2}$, which is one to two orders of magnitude higher than found for previous analyses of data taken at similar orbital phases. Such a large column produces a Compton shoulder in the Fe K$\alpha$ line. The transition appears to be caused by the onset of efficient cooling, which cools the plasma by 10 million degrees in just 10 ks, allowing matter to enter the magnetosphere. This happens after a major disturbance, probably the arrival of a train of wind clumps with individual masses in the range $10^{19-20}$ g. This train moves ballistically in an eccentric orbit around the NS, producing a distinctive Doppler modulation in the \ion{Fe}{xxv} line.}


\titlerunning{State transition of Cen X-3 with Chandra}

\authorrunning{Sanjurjo-Ferr\'{\i}n et al. }

\maketitle



\section{Introduction}

high-mass X-ray binaries (HMXRBs) are binary stellar systems characterized by a compact object, either a neutron star (NS) or a black hole, orbiting a massive O or B-type star known as the companion. The X-ray radiation in these systems is generated through the accretion of matter from the companion's powerful stellar wind, typical of these early type stars, onto the compact object. Consequently, these systems serve as valuable laboratories for the study of accretion processes, accretion stream structures, and the properties of the companion's stellar wind \citep[][for a review]{2017SSRv..212...59M}.

Cen X-3 was first observed by \citet{641ee8007ab8411c818bcc484f398d2f}. This is an eclipsing binary system composed of a NS and its giant O6-8 III counterpart V779 Cen \citep{1972ApJ...172L..79S, 1974ApJ...192L.135K, 1979ApJ...229.1079H}. The NS remains eclipsed, that is occulted behind the donor star with respect to the observer, $\sim$ 20\% of the orbit. The NS spin period of 4.8~s was discovered with the \textit{Uhuru} telescope \citep{1972ApJ...172L..79S,1971ApJ...167L..67G}.

Cen X-3 has a rather short orbital period of only 2.1 days, which allows a wide range of orbital phases to be studied in a single focused observation. The optical star fills its Roche lobe, indicating the probable formation of an accretion disk \citep{1986A&A...154...77T,1978ApJ...224..625P} in agreement with the high luminosity of the source, the presence of quasi-periodic oscillations (QPOs) at $40$ mHz \citep{1991PASJ...43L..43T,2008ApJ...685.1109R}, and the observed trend of the NS's spin-up at about $1.135$ milliseconds per year \citep[albeit with some fluctuations;][]{1996ApJ...456..316T}.

\begin{figure}
\includegraphics[trim={0cm 0cm 0cm 0cm},width=1\columnwidth]{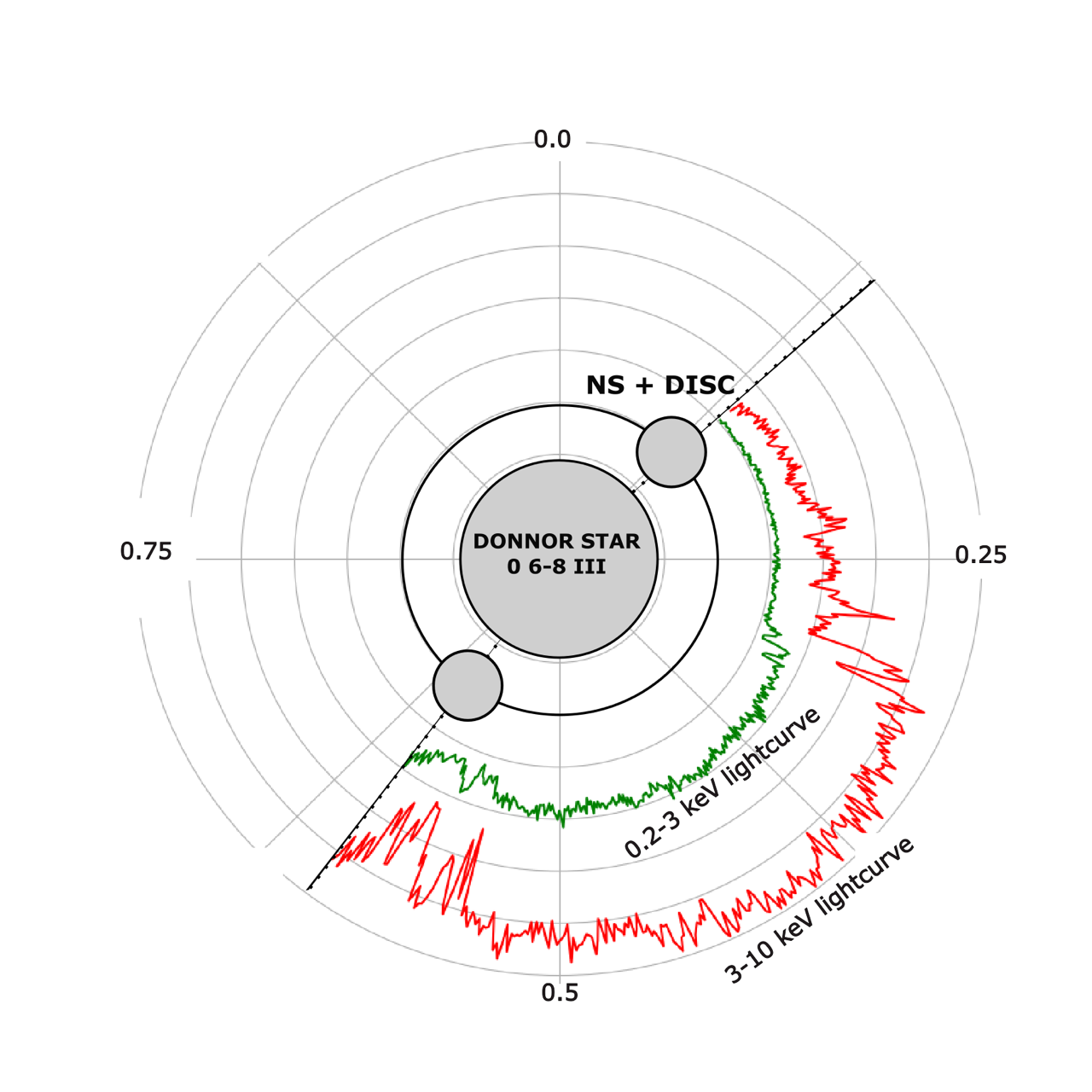}
\caption{Pole-on scaled sketch of the system. The light curve obtained during this observation is plotted along the orbital phases covered. The radius of the donor star, the orbit and the accretion disk are to scale. Phase $\phi=0.0$ corresponds to the NS mid eclipse and phase $\phi=0.5$ to the NS inferior conjunction.} 
\label{sketch}
\end{figure}

\begin{table*}
\caption{Properties of the Cen X-3 System}
\centering
\begin{tabular}{lcc}
\midrule\midrule\\
\multicolumn{3}{c}{Donor Star}\\
\midrule
    MK type & O6-8 III & \citet{1979ApJ...229.1079H}\\
    $M_{\rm opt} $ & $20.2^{+1.8}_{-1.5}$  $M_\odot$ & \citet{2007AA...473..523V} \\
    
    $R_\ast$ & $12.1 \pm 0.5 ~ {R_\odot}$ & \citet{2011ApJ...737...79N} \\
        $E(B-V)$  &       2.456 {\rm mag}             & \citet{2013MNRAS.431..394W}\\
\midrule          
\multicolumn{3}{c}{Neutron Star}\\
\midrule
     $M_{\rm NS}$ & $1.34^{+0.16}_{-0.14}$  $M_\odot$  & \citet{2007AA...473..523V} \\
    Spin period & 4.79 s & This work \\
   
     Magnetic field    & $(2.4-3.0) \times 10^{12}$  G &  \citet{2011ApJ...737...79N}\\
\midrule            
\multicolumn{3}{c}{Orbit}\\ 
\midrule
    Orbital period  & 2.087139842(18) d  & \citet{2023}\\
    $\dot P $ & {-}1.03788(27)  $\times10^{-8}$  days day$^{-1}$ & \citet{2023}\\
    $i$ & $72$\degree $^{+6}_{-5} $  & \citet{falanga} \\
    Eccentricity $e$ & 0.00006 - 0.000266  & \citet{2023} \\
    Semimajor axis $a$   $1.58^{ +0.05}_{-0.04}$  ~ $R_\ast$ & \citet{2021MNRAS.501.5892S}\\
    NS distance to the barycenter & $1.48 \pm 0.01$  ~ $R_\ast$ & \citet{2021MNRAS.501.5892S}\\
    NS velocity with respect to the barycenter & $436 \pm 4$ km s$^{-1}$  & \citet{2021MNRAS.501.5892S}\\
    Distance & $6.8^{+0.6}_{-0.5}  ~ {\rm kpc}$ & \citet{2021yCat.1352....0B}\\
    $T_{0}$ (MJD) &  50506.788423(7)  &  \citet{falanga}  \\
  \hline
\end{tabular}

\label{parameters}

\end{table*}



Periodic high and low X-ray states on timescales of about $125$ to $165$ days have been reported \citep{1983ApJ...273..709P}. In addition, \citet{2022RMxAA..58..355T} reported the presence of a super-orbital period of $220 \pm 5$ days, attributed to the precession of the accretion disk. There may be other structures contributing to these modulations, such as an accretion wake \citep{Suchy_2008}, making Cen X-3 one of the most complex binaries known.

\citet{2010RAA....10.1127D} note that in high-soft states the eclipse egress and ingress are sharper and shorter while in low-hard states they are shallower and longer, concluding that different flux states are caused by a varying degree of obscuration by the precessing accretion disk. In \cite{2024MNRAS.529L.130L} a link between torque reversals and the orbital profile is suggested.

A cyclotron resonance scattering feature (CRSF) at approximately 30 keV was initially detected with \textit{Ginga} \citep{1992ApJ...396..147N} and confirmed by \textit{BeppoSAX} \citep{1998A&A...340L..55S} and \textit{RXTE/HEXTE} \citep{1999hxra.conf...25H}. Additionally, \cite{article} observed that the CRSF energy decreases along the NS pulse, starting at 36 keV during the ascent and reaching 28 keV during the descent. This phenomenon can be explained by assuming an offset of the dipolar magnetic field relative to the NS center. Furthermore, \cite{2023MNRAS.519.5402Y} reported the presence of two CRSFs, with the fundamental line at approximately 28 keV and the harmonic line at approximately 47 keV.

The stellar wind of the donor is partially photo-ionized by the X-ray emission, producing emission lines due to recombination \citep{1996ApJ...457..397A,2003ApJ...582..959W}. These line fluxes change with the orbital phase and their equivalent widths ($EW$) are especially enhanced during the eclipse, when the direct continuum produced by the NS is blocked by the optical counterpart. 

The analysis presented in this paper heavily relies on the findings previously reported in \citet{2021MNRAS.501.5892S} (Paper 1 from now on). In this previous work, two \textit{XMM-Newton} observations were analyzed, one taken out of eclipse in a high-soft state (orbital phases $\sim \phi=0.35-0.8$) and the other during the eclipse-egress in a low-hard state (orbital phases $\sim \phi=0.0-0.36$). The main results  of Paper 1 are summarized below:

\begin{enumerate}
\item A spectral model consisting of a \texttt{powerlaw + bbody + $\sum_{i=1}^{17}$ Gaussian$_{i}$}, where the Gaussian functions account for the emission lines, describes the observed spectra well. During the high-soft state, the blackbody component increased in the total emission budget, although the power law dominated the entire spectrum at all fluxes. The blackbody emission area had an equivalent radius of the order of 2 km, compatible with a hot spot on the NS surface. During low states, the size of the blackbody emitting area was found to be much larger, ranging from 5 to 10 km.

\item The absorption to the X-ray source was modeled as $(C\exp(-N_{\rm H,1}\sigma(E)) +(1-C)\exp\left(-N_{\rm H,2}\sigma(E) \right))$, where $C$ acts as a covering fraction to account for the clumpiness of the stellar wind and can vary from 0 to 1. $N_{\rm H,1}$ was compatible with the interstellar medium (ISM), while $N_{\rm H,2}$ is consistent with being mostly local\footnote{Modeled in this manner}, $N_{\rm H,2}$ also includes the more modest contributions from the ISM. See Eq. \ref{model} for an alternative definition of the partial covering., decreased as the egress progressed, from $(7-8)\times 10^{22}$ cm$^{-2}$  to $\sim 4\times 10^{22}$ cm$^{-2}$ out-of-eclipse.

\item The spectra exhibited emission lines likely originating from a photoionized plasma. The equivalent widths of highly ionized species (\ion{Fe}{xxv} He-like and \ion{Fe}{xxvi} H-like Ly$\alpha$) decreased during the egress as the continuum rose. Conversely, their fluxes, measured through the areas of the Gaussian functions, increased during the egress, indicating that a significant fraction were originated relatively close to the NS. Additionally, the equivalent width of neutral Fe increased, suggesting that the line originates in dense and cold structures emerging during the egress. In Paper 1 we proposed that these structures are located along the accretion stream.  In this scenario, the line fluxes should exhibit maximum and minimum values at orbital phases 0.25 and 0.5, corresponding to maximum and minimum projected areas of the accretion stream. 

\item The low-hard state egress $(3-10)$ keV light
curve was highly structured, showing several intervals. These intervals corresponded with structures emerging from the egress. The first interval corresponded to a structure whose size was compatible with the Roche lobe size of Cen X-3, $R_{\rm L}\simeq 0.3R_{*}$. The second rise signaled the egress of an emitting region with a size $\simeq 0.17R_{*}$. Possible candidates for these structures could be the accretion disk and reflection from the optical star atmosphere close to the inner Lagrangian point, respectively.

\item Some deep and long dips (with durations on the order of 1000 s and an average flux drop of 40\%) could be observed in the out-of-eclipse (high-soft state) light curve. They could be caused by a sporadic decrease in the accretion rate, most likely due to instabilities at the inner edge of the disk interacting with the NS magnetosphere. The characteristic time for these instabilities scales with the diffusion time at the inner disk radius, $t_{\rm d}\sim (1.0-1.5)\times 10^{3}$ s for Cen X-3, close to the observed dip duration.


\end{enumerate}

The objective of this article is to analyze the target of opportunity observation (ToO) obtained in the low-hard state, focusing on the egress phase in order to confirm the presence of emerging structures during the eclipse. Additionally, phase 0.5 was targeted to examine the behavior of Fe lines in a low-hard state outside the eclipse. 

To ensure the observation was timed correctly, a machine learning (ML) Random Forest algorithm (formed by collection of decision trees combining their predictions to achieve more accurate and stable results) was developed. The data used to classify the low and high states was the Monitor of All-sky X-ray Image (MAXI); \citealt{2009PASJ...61..999M}) orbital light curve. The observation was made only 9 days after the trigger (Fig. \ref{maxi}).

In Table \ref{parameters} we compile the system parameters relevant for this work, and in Fig. \ref{sketch} we show a sketch of the system where the orbit, the donor, and the putative disk (the size deduced from this work) are to scale with phase $\phi=0$ corresponding to mid eclipse.

The paper is organized as follows. In \S2 we describe the observations and the analysis performed. In \S3 we present our results. In \S4 we discuss the results obtained and in \S5 we summarize our findings.

\section{Observations and analysis}
\label{sec:obs}

We have analyzed a pointed (ToO) observation performed with the High-Energy Transmission Grating spectrometer \textsc{hetg} \citep{Canizares_2005} on board the \textit{Chandra} X-ray observatory. There are two sets of gratings available: the High-Energy Grating (\textsc{heg}) offering a resolution of 0.011 \AA $ $  in the bandpass of approximately 1.5 to 16 \AA\ , and the Medium-Energy Grating (\textsc{meg}) offering a resolution of 0.021 \AA\ in the bandpass of approximately 1.8 to 31 \AA. In this observation the \textsc{heg} and \textsc{meg} spectrometers acquired data uninterruptedly for 83.89 ks. The log of observations is presented in Table\ref{chobs}.

\begin{figure}
\includegraphics[trim={0cm 0cm 0cm 0cm},width=1\columnwidth]{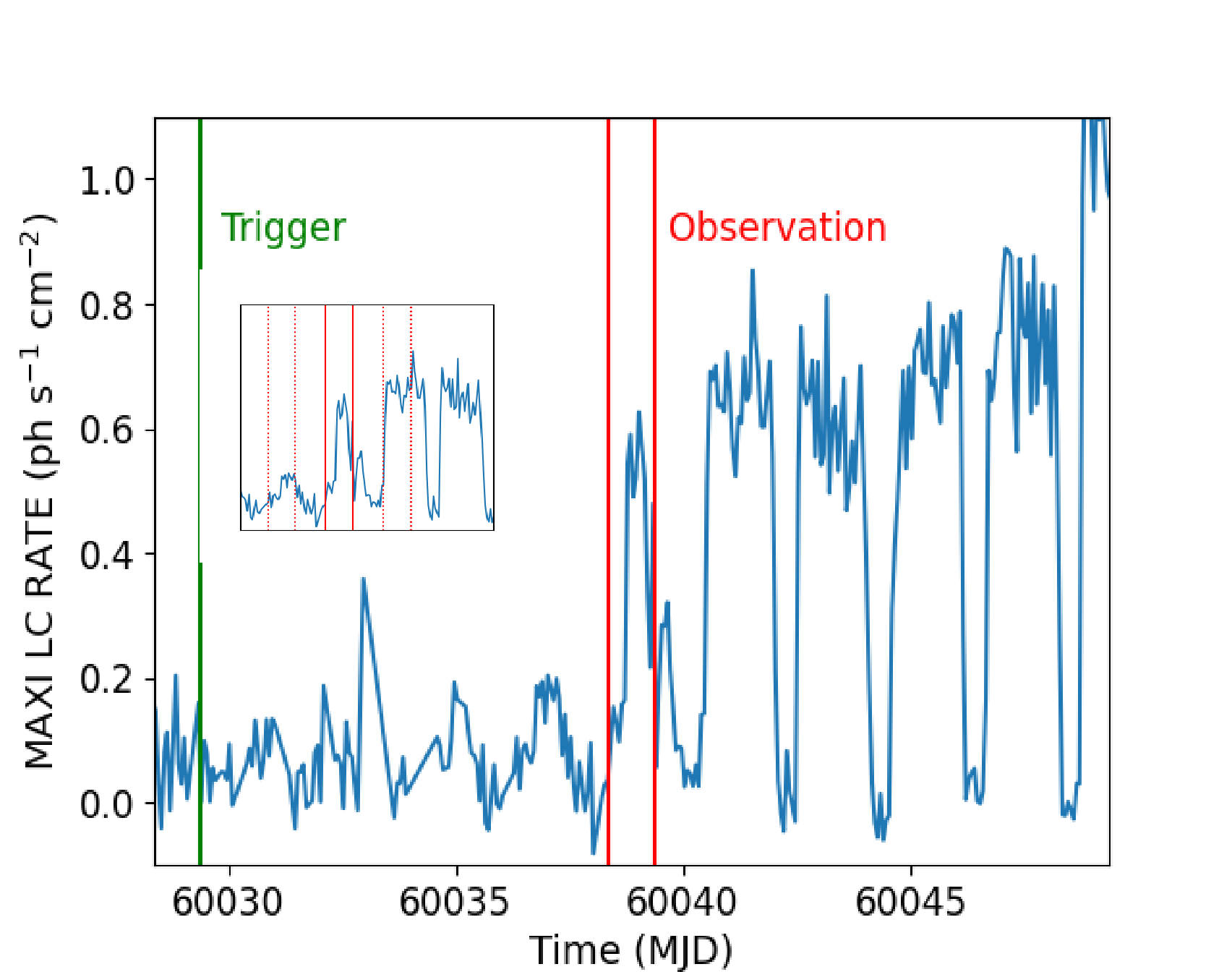}
\caption{Long-term Cen X-3 MAXI light curve. The green vertical line represents the trigger, that is, the moment when our ML algorithm gave us a signal with a high probability of a long low state. The observation between the two red solid lines was made 9 d later, coinciding with the start of the source's state change. The inset is an enlargement of our observation (marked between two red solid lines) and the same phases one orbit before and after (marked between two red dashed lines). The light curve shown in Fig. \ref{sketch} corresponds to the part of this light curve between the two red solid lines. Note that the MAXI light curve covers the $2-20$ keV energy range, while the light curves shown in Fig. \ref{sketch} cover the $0.2-3$ (green) and $3-10$ (red) keV energy ranges, respectively, and are presented in polar coordinates. } 
\label{maxi}
\end{figure}

Although the source was in a low state at the beginning of the observation, it suddenly changed from a low-hard to a high-soft state about a third into the observation. Although this fact prevented us from performing the desired experiment, we had the unique opportunity to observe the source during the state transition at high spectral resolution. This transition was sudden and sharp, occurring around phase $\phi \sim$ 0.27, although in later phases (from phases $\phi \sim$ 0.48 onwards) the count rate decreased again. However, the following orbits showed a real recovery to a high soft state. This phenomenon is clearly seen in Figure \ref{maxi} (see inset) when we compare the light curve for the same orbital phases but in the subsequent orbit when the system is fully in the high state (dotted red lines to the right of our observation). The beginning of our observation coincides with the exact moment of the sharp egress, while the end of our observation occurs well before the ingress of the eclipse.

\begin{table}
\caption {\emph{Chandra} Observation log. }
\centering
\begin{adjustbox}{max width=\columnwidth}
\begin{tabular}{cccc}
\hline\hline
Observation ID & Date & Orbital phase & GTI Duration \\
  &    &  &  (ks)\\
\midrule
   26512 &  2023.04.04 08:50   &   0.13-0.59 &  83.89  \\
  \hline
\end{tabular}
\end{adjustbox}
\label{chobs}
\end{table}

The spectra and response (arf and rmf) files were generated using standard procedures with the \textsc{ciao} software (v4.15, CalDB 4.15). First dispersion orders ($m = \pm 1$) from \textsc{heg} and \textsc{meg} were extracted and combined for each of the phase-resolved spectra analyzed in this article. Useful data were considered in the $(0.2-10)$ keV
range. The spectra were analyzed and modeled with the \textsc{isis}\footnote{https://space.mit.edu/cxc/isis/} package. The emission lines were identified thanks to the \textsc{atomdb}\footnote{http://www.atomdb.org/} data base.

\begin{figure*}
\centering
\subfigure{\includegraphics[trim={4cm 2cm 4cm 4cm},width=0.9\textwidth]{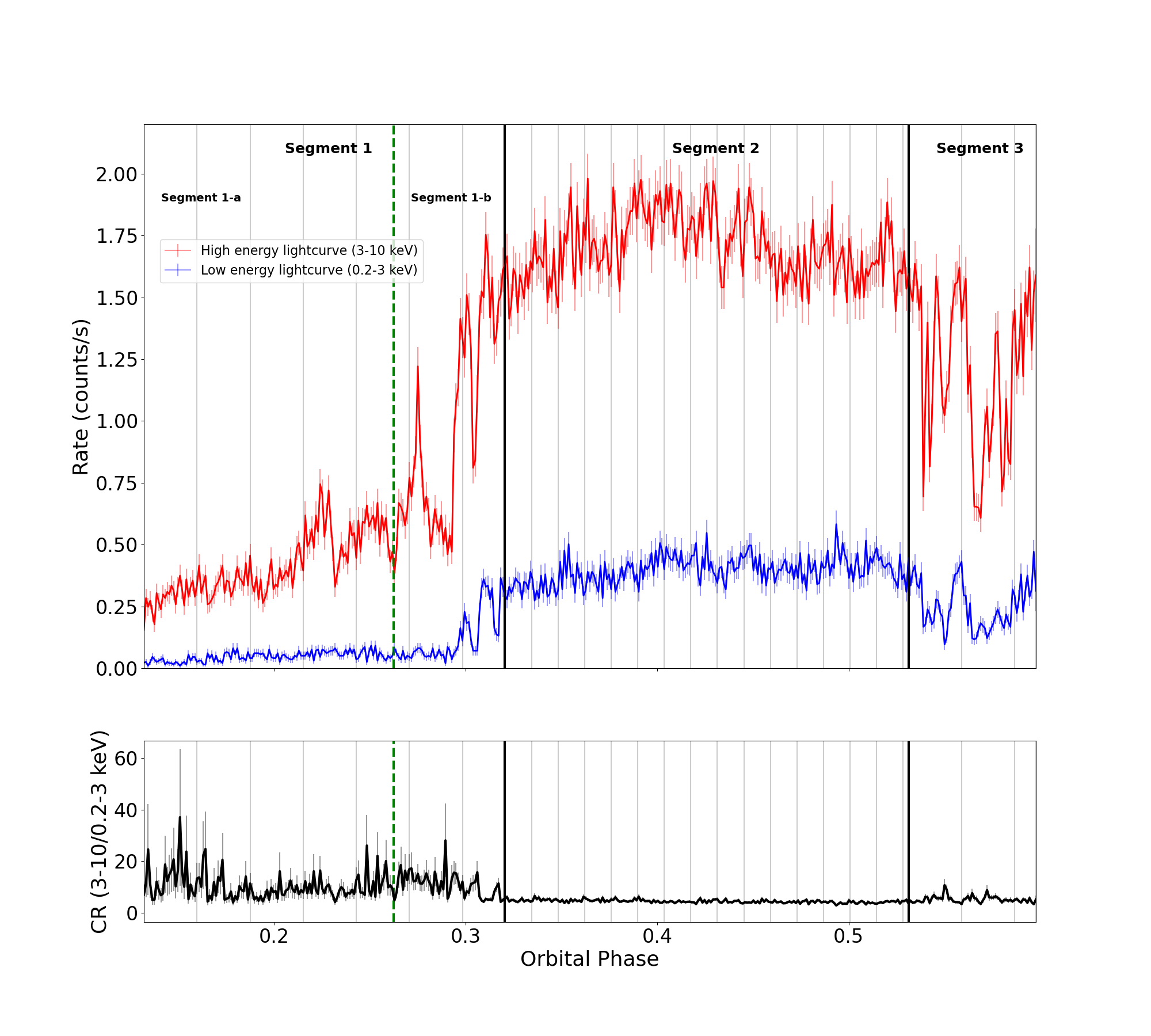}}

\caption{ \textit{Chandra} Cen X-3 200s bin X-ray light curve. The top panel shows the high energy light curve: $3-10$ keV (red) and the low energy light curve: $0.2-3$ keV (blue) respectively. The bottom panel shows the color ratio (CR, black) defined as ($3-10$) keV/($0.2-3$) keV.} 
\label{lcurve}
\end{figure*}



\section{Results}
\label{res:intro}

To perform the spectral and temporal analysis, the observation was divided into three main segments (hereafter referred to as segments 1, 2 and 3). Segment 1 was further divided into segment 1-a (low state) and segment 1-b (rise). These subdivisions reflect a very clear change in the behavior of the high and low energy light curves and thus their color ratio (CR; Fig. \ref{lcurve}). Each of these segments has been divided into intervals. Segments 1-a, 1-b and 3 were divided into 5000 s intervals and segment 2 into 2500 s intervals in order to accommodate the higher count rate which allows for more detailed analysis.

\subsection{\textit{Chandra} timing}
\label{res:sec:timming}

\subsubsection{Pulse evolution}

To investigate the evolution of the NS spin pulse caused by the Doppler effect as it travels along its inclined orbit, we divided 
the light curve into blocks of 3000 bins each (i.e. 600 s), using a sliding window approach with a 100 bin step (20 s). In a sliding window a fixed-size segment of data moves step-by-step across a dataset. This technique allows for detailed, overlapping analysis of the entire dataset.

In each of these blocks a \textit{Lomb-Scargle} periodogram \citep{1982ApJ...263..835S} was applied. From all the results obtained we selected  as good results the ones with periods between 4.6 and 5.0 s and a false alarm probability of less than 10$^{-5}$. 
The same procedure was used in the \textit{XMM-Newton} out of eclipse light curve in order to compare results. 
For the \textit{Chandra} observations, 4261 sections were analyzed, with only 131 deemed to be good. In contrast, for the \textit{XMM-Newton} observations, 3949 sections were analyzed, and 3934 were considered good.

The NS pulse Doppler evolution caused by the NS travel through the orbit can help us to constrain the orbital parameters. Ideally, this would require simultaneous radial velocity curves for the optical donor. Since this is not available here, we use {\it Chandra} data to constrain orbital values. For a pulsar with period $P_0$ on an orbit with an inclination angle $i$ and a radial velocity, $v_{\rm D}$, of the NS with respect to the system barycenter, the observed Doppler-shifted spin period, $P_{\rm D}$ is,

\begin{equation}
\label{Doppler}
\begin{split}
P_{\rm D}=P_{0}\left(1+\frac{v_{\rm D}}{c}\right) ~~,~~\\
v_{\rm D}=-r\omega \sin\phi \sin i\\
\end{split}
\end{equation}

\noindent where $\omega$ is the angular velocity and $r$ is the orbital radius, that is the distance from the NS to the system barycenter. 

The model was fitted using a \texttt{particle swarm optimization} algorithm (PSO). In PSO each particle represents a potential solution to an optimization problem. The particles move through the solution space to find the optimal parameter configuration. This approach yields slightly different results for each run due to its random nature. The provided errors correspond to the standard deviation of the obtained results \citep{10.1162/EVCO_r_00180}. The results are collected in {Table \ref{orbital_parameters} (see also Figure \ref{doppler}).

\begin{table}
\caption {Cen X-3 orbital parameters derived from the NS's spin pulse Doppler shifts observed in \textit{XMM-Newton} and \textit{Chandra} data.}
\centering
\begin{adjustbox}{max width=\columnwidth}
\begin{tabular}{ccc}
\hline
Parameter &  XMM-Newton & Chandra \\
\hline

Semimajor ($R_\star$) & $1.6 \pm 0.1$ & $1.6 \pm 0.1$ \\
Inclination (deg) & $73.0 \pm 2.6$ & $74 \pm 3$ \\
$R_\star$ ($R_\odot$) & $12.0 \pm 0.2$ & $12.0 \pm 0.2$ \\
$M_{\rm opt}$ ($M_\odot$) & $21.0 \pm 0.8$ & $22 \pm 1$ \\
$M_{\rm NS}$ ($M_\odot$) & $1.34 \pm 0.10$ & $1.35 \pm 0.10$ \\
NS spin period (s) & $4.80500 \pm 0.00013$ & $4.7940 \pm 0.0004$ \\

\hline
\end{tabular}
\end{adjustbox}
\label{orbital_parameters}
\tablefoot{The fits yielded $r^2$ values of approximately 0.9 for both datasets.}
\end{table}

\begin{figure}
\includegraphics[trim={0cm 0cm 0cm 0cm},width=1\columnwidth]{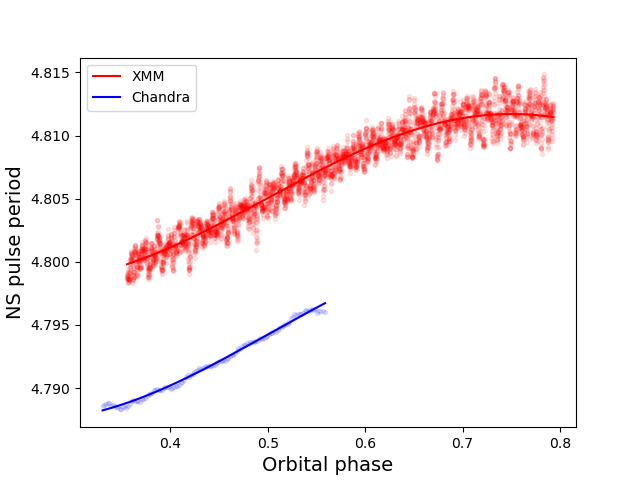}
\caption{Observed NS spin period evolution. Blue and red represent the \textit{Chandra} (2023)  and  the \textit{XMM-Newton} (2006) NS spin evolution. Red  and blue lines represent the theoretical Doppler evolution calculated with the orbital parameters deduced from the observed data for each observation (see Table \ref{orbital_parameters}).}
\label{doppler}
\end{figure}

\subsubsection{Dips}

Some deep and long dips (with durations of the order of 1000 s and an average flux drop of 40\%) could be observed in the high state {\it XMM-Newton} observation (Paper 1). These dips were attributed to the interaction of the NS magnetosphere with the inner edge of the accretion disk (Paper 1, its Fig. 3).

We looked for such features in current \textit{Chandra} observation, but they are not so obvious as in the {\it XMM-Newton} light curve, where they could be easily seen by eye. Thus, we developed an automatic algorithm that detects dips and collects their main characteristics and applied it consistently to both the \textit{Chandra} and \textit{XMM-Newton} light curves. 

Dips present in HMXBs light curves can be caused by different astrophysical phenomena, as for example clumps in the stellar wind, inter-clump medium, decrease of the accretion rate, interactions of the NS magnetosphere with an accretion disk, etc.

To ensure the validity of these dips, 
we first require that dips are consistently present in both high-energy ($3-10$ keV) and low-energy ($0.2-3$ keV) light curves, with overlapping duration of at least 60\%. To distinguish genuine dips from noise-induced artifacts, we generate artificial noise through de-trending and shuffling of the original light curves. Dips that align between the original data and noise-generated data are considered false positives and are subsequently eliminated. This systematic approach guarantees the reliability of our dip identifications.

We define the depth of the dip as the decrease in rate divided by the height of the contour lines at which the widths of the dips were evaluated. Contour lines connect points where the light curve count rate has the same value. We define the depth ratio as the depth of the dip in the low energy region divided by the depth of the dip in the high energy region. A depth ratio greater than 1 means that the dip is more pronounced in the low energy region than in the high energy region. Since low energy radiation (below 3 keV) is easily absorbed by the matter it passes through, dips with a depth ratio greater than 1 suggest an absorption origin and are therefore considered clump candidates. Conversely, dips with a depth ratio less than 1 are more pronounced in the high energy range than in the low energy range, indicating that they are likely caused by an intrinsic decrease in accretion power due to irregularities in the stellar wind.
The dips are shown in Fig. \ref{dip_orbit}, plotted against the light curve (light grey). As can be seen, a series of dips compatible with clump signatures is present during and after the state transition (end of segment 1 and on).

\begin{figure}
\includegraphics[trim={-2cm 3cm 2cm 0cm},width=0.8\columnwidth]{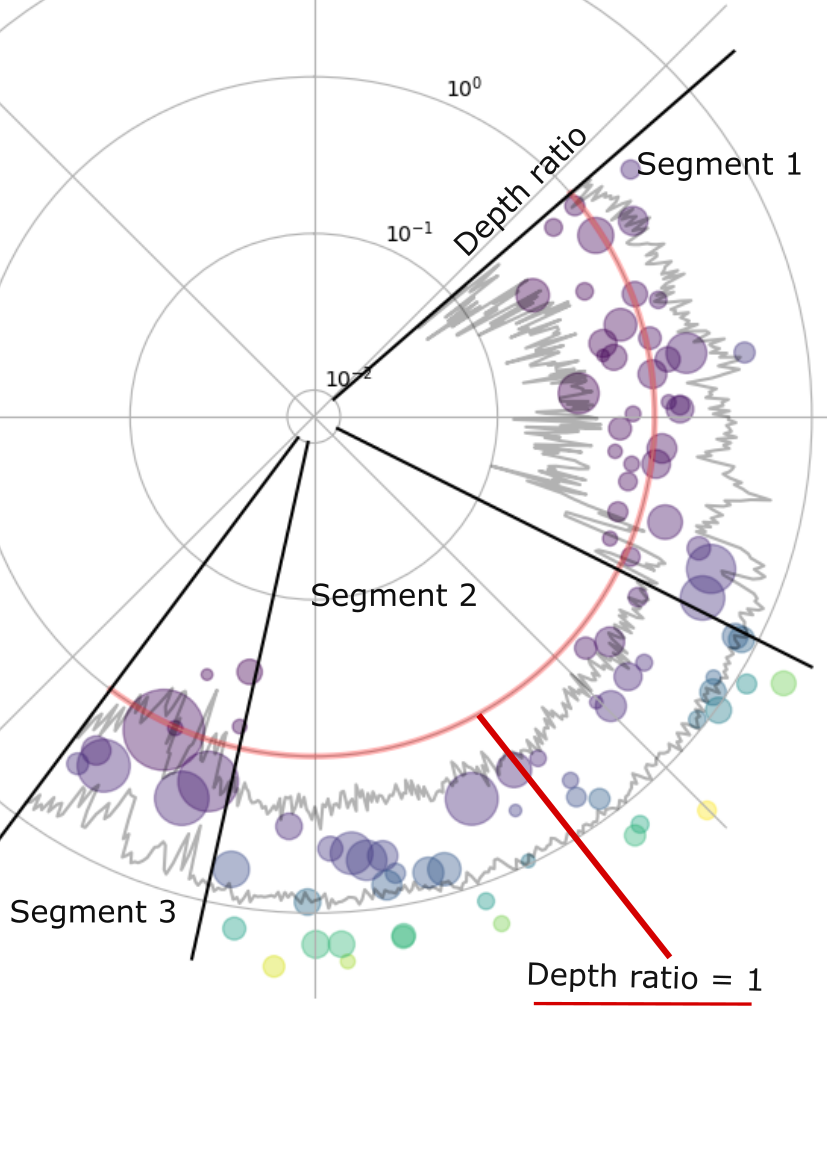}
\caption{Dips detected during this observation. The radial Y axis represents the depth ratio while the dot size is proportional to the dip duration. The color gradient from blue to yellow in a crescent pattern serves to accentuate the relative depth of each dip. A depth ratio higher than one is indicative of absorption (clump signatures) while depth ratio equal or smaller than one is indicative of intrinsic decrease of the flux. Note the logarithmic scale in the radial y axis.} 
\label{dip_orbit}
\end{figure}

Once we have the dip set for each observation, we apply \textit{kernel density estimation} (KDE) to the result. KDE produces a smooth curve (probability density function) that represents the underlying distribution of the data. It does this by placing a kernel (a smooth, symmetric and usually bell-shaped function) at each data point and summing these kernels to produce the final smooth curve. It should be noted that the efficiency of detecting the dips, and the results themselves, are affected by the time resolution of each light curve, and that we are comparing results from two different instruments.

\begin{figure*}
\includegraphics[trim={0cm 0cm 0cm 0cm},width=1\textwidth]{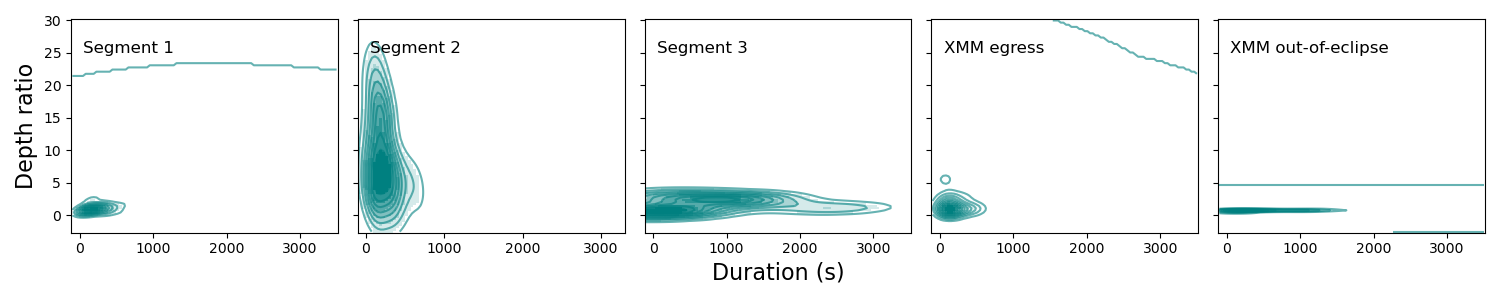}
\caption{Kernel density estimation (KDE) plots depicting the relationship between 'Depth Ratio' (low energy depth divided by high energy depth) and dip duration in various observations. The panels showcase results from distinct datasets. From left to right: \textit{Chandra} segments 1, 2 and 3, \textit{XMM} high-soft observation and \textit{XMM} low-hard observation results.
Solid lines within each panel represent KDE contour levels, demarcating regions of equal density in the estimated distribution.}
\label{dips}
\end{figure*}

In Fig. \ref{dips} we represent the underlying distribution of the duration vs the depth ratio for the dips detected in each observation. 
Segment 1 shows a similar distribution to the {\it XMM-Newton} low-hard observation: they show a short duration with a tight range of depth ratios. In the {\it XMM-Newton} high-soft observation the dip distribution shows a narrow and close to 1 depth ratio but with a longer duration, compatible with dips produced by the interaction of the NS magnetosphere with the inner edge of the accretion disk. In Segment 2, dips show a wide variety of density ratios, pointing to a higher clump density. During Segment 3 dips are longer than those observed in Segments 1 and 2, but still show a wide range of depth ratios.

In order to ascertain the overall incidence of absorption throughout the observation, we calculated the sum of the contributions from each clump candidate. This contribution is determined by its depth ratio (proportional to its absorbing power) weighted by its duration (proportional to its size). This value is subsequently divided by the total duration of the corresponding section. The result is shown in Fig. \ref{den_sec}. A larger value of this metric indicates a more significant presence of absorbing material within the interval.
 
\begin{figure}
\includegraphics[trim={0cm 0cm 0cm 0cm},width=\columnwidth]{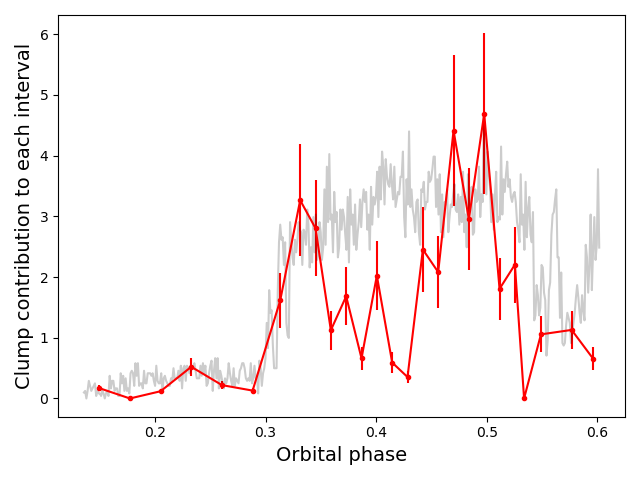}
\caption{Clump contribution to each interval (red, superimposed on the light curve). It indicates the impact of absorbing material within each interval through the presence of clump signatures.}
\label{den_sec}
\end{figure}
%
\subsection{\textit{Chandra} spectra}
\label{res:sec:spec}

To observe the evolution of the parameters throughout the orbit, we performed a phase-resolved spectral analysis, obtaining a spectrum for each of the intervals defined in Fig. \ref{lcurve} (clear grey lines). The same spectral model was used to describe all the spectra. Following paper 1, and in order to compare our results with it, our first attempt was to combine a black body (\texttt{bbody}) plus a power law (\texttt{powerlaw}) modified by absorption, but the parameters did not always converge and the \texttt{powerlaw} contribution was always negligible. Instead, all spectra could be satisfactorily fitted with a single absorbed black body. 

The \texttt{bbody} model parameters include the temperature ($kT_{\rm bb}$) and the normalization, defined as $L_{39} D_{10}^{-2}$ where $D_{10}$ is the distance to the source in units of 10 kpc and $L_{39}$ is the luminosity in units of $10^{39}$erg s$^{-1}$. Besides the interstellar medium (ISM) absorption component, we also allowed for the presence of local absorption, modulated by a partial covering fraction (parameter \textit{C}) which acts as a proxy for the degree of clumping in the stellar wind of the donor star. The ISM absorption is modeled by the X-ray absorption model 
T\"{u}bingen-Boulder \texttt{TBnew}\footnote{http://pulsar.sternwarte.uni-erlangen.de/wilms/research/tbabs/}. This model calculates the cross section for X-ray absorption by the ISM as the sum of the cross sections due to the gas-phase, the grain-phase, and the molecules in the ISM \citep{2000ApJ...542..914W}. 

The model used is described by Eq. \ref{model}.

\begin{equation}
\label{model}
\begin{split}
F(E) &= \left[ C \exp\left(-N_{\rm H}^{\rm LOC} \sigma(E)\right) + \exp\left(-N_{\rm H}^{\rm ISM} \sigma(E)\right) \right] \\
& \quad \times \left[ \text{bbody} + \sum_{i=1}^{16} \text{Gaussian}_{i} \right]
\end{split}
\end{equation}

Where the summation represents the emission lines present within the spectra. Each emission line was fitted with a single Gaussian function, added to the \texttt{bbody} component representing the continuum, all of them modified by the absorption component. The interstellar absorption $N_{\rm H}^{\rm ISM}$ was fixed to the ISM absorption towards the optical companion ($1.3\times 10^{22}$ cm$^{-2}$) using $E(B-V)$ from Table \ref{parameters} \citep{2015ApJ...809...66V}.

The transition from low to high state is sharp and sudden. This transition is clearly not due to the eclipse exit, since it occurs at a rather advanced orbital phase $\sim 0.27$ (see Fig. \ref{maxi}). This can be seen not only in the light curve (transition from segment 1 to segment 2, Fig. \ref{lcurve}) but also in the spectra (Fig. \ref{all_spectra_spec}), which change completely from segment 1-a (red) to segment 1-b (green).

\begin{figure*}
\sidecaption
\includegraphics[trim={0cm 0.5cm 0cm 0.5cm},width=12cm]{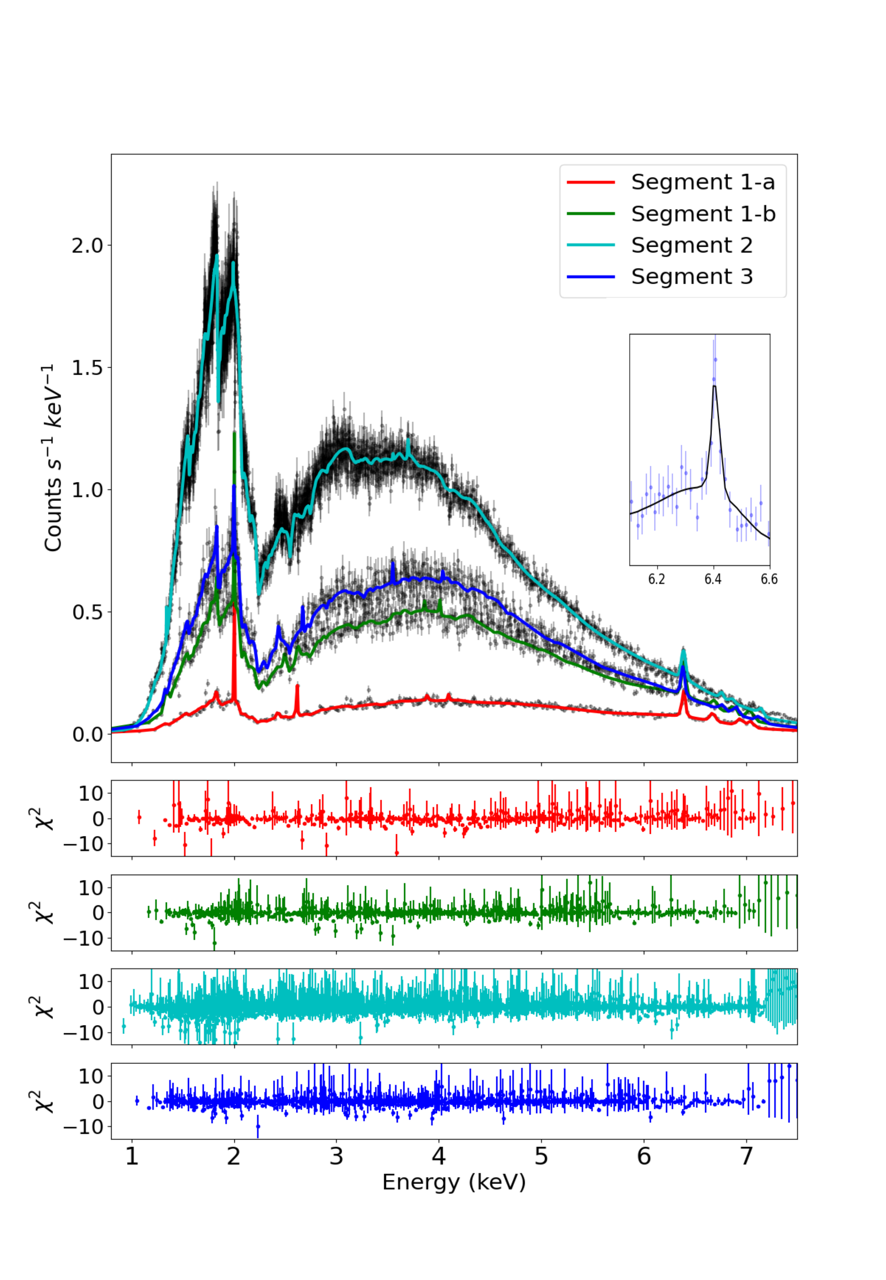}
\caption{Spectra of the different segments as described in Fig \ref{lcurve}: Segment 1-a (red), Segment 1-b (green), Segment 2 (cyan) and Segment 3 (blue). The four lower panels represent $\chi^2$ of each of the models. The model parameters are collected on Table \ref{parameters}.}
\label{all_spectra_spec}
\end{figure*}

\begin{figure*}
\centering
\subfigure{\includegraphics[trim={0cm 0cm 0cm 0cm},width=1\textwidth]{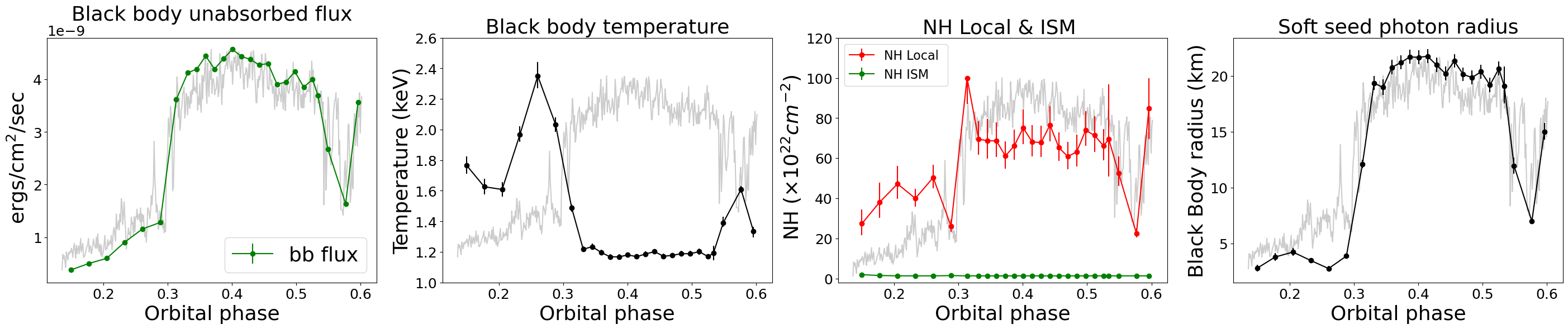}}
\caption{Evolution of the main model parameters for each interval, as described in Fig. \ref{lcurve}, superimposed to the light curve (light grey). From right to left: black body flux, corrected for absorption, the black body temperature, the interstellar (ISM) (green) plus the local (red) absorption and the soft seed photon radius.} 
\label{all_spectra_par}
\end{figure*}

\begin{table*}
  \centering
  \caption{Best-fit model parameters for each of the different four segments }
  \begin{adjustbox}{max width=\textwidth}
    \begin{tabular}{ccccccccc}
      \hline\hline
      \\
      Spectra & $ \chi_{\rm r}^2$ & $N_{\rm H}^{\rm LOC}$ & $C$ & $N_{\rm H}^{\rm ISM}$ &  $K_{\rm bb}$ & $kT_{\rm bb}$ & $F^{\rm unabs}_{\rm bb}$ & $R_{\rm W}$ \\\\

     &&($\times10^{22}$ cm$^{-2}$)& & ($\times10^{22}$ cm$^{-2}$)& ($\times$10$^{-3}$) &(keV) &  ($\times$10$^{-10}$ergs cm$^{-2}$ s$^{-1}$)& (km) \\
     
      \midrule

Segment 1-a 	&	1.6	&	56 $^{+5}_{-4}$  & 1.000    $\pm$ 0.014 &1.300 $\pm$0.008                 &6.5 $\pm$ 0.3              &2.10 $\pm$ 0.04       &6.9 $\pm$ 0.3      &2.700 $\pm$ 0.003\\\\
Segment 1-b 	&	1.2	&	48 $^{+5}_{-4}$  & 1.00     $\pm$ 0.04     &1.30 $\pm$ 0.03                     &14.0 $^{+0.5}_{-0.4}$   &1.60 $\pm$ 0.02      &22.0 $\pm$ 0.2    &8.700 $^{+0.008}_{-0.009}$\\\\
Segment 2 	    &   1.6	&	75.0 $^{+2.0}_{-1.9}      $  & 1.000 $\pm$ 0.003   &1.300 $\pm$ 0.001   &27.00 $\pm$ 0.18       &1.200 $\pm$ 0.005    &43.00 $\pm$ 0.04 &20.00 $\pm$ 0.02 \\\\
Segment 3 	    &	1.0	&	55 $^{+7}_{-6}$ & 0.85     $^{+0.12}_{-0.11}  $  &1.300 $\pm$ 0.015   &18.00 $\pm$ 0.08       &1.50 $\pm$ 0.03      &23.00 $\pm$ 0.03  &9.600 $^{+0.009}_{-0.010}$\\\\

      \hline
    \end{tabular}
  \end{adjustbox}
  \label{table:model_parameters}
  \tablefoot{See Fig. \ref{lcurve} for segment division and Fig. \ref{all_spectra_spec} for the evolution of the different model parameters through the observation. The best-fit model parameters obtained for each interval are collected in the Appendix Table \ref{appendix:A}.}
\end{table*}

\subsubsection{Continuum}
The results of the fits for each segment are presented in Table \ref{table:model_parameters}. The spectral model parameters of each interval are collected in the Appendix, Table \ref{Appendix:A}. In Fig. \ref{all_spectra_par}, lower panel, the orbital evolution of some important parameters is superimposed on the light curve (light grey). 

The overall contribution of the black body closely follows the light curve (left panel). 
But drastic changes occur during the brightness rise (Segment 1-b): The temperature of the blackbody suddenly drops to the lowest values in approximately 10 ks, as shown in the second panel. The onset of this cooling ($\phi\sim 0.27$) precedes the onset of the state change (flux increase, $\phi \sim 0.30$), moment at which the local absorption experiences a moderate rise. The radius of the soft photons production site also increases from $(2-5)$ km in Segment 1 to approximately 20 km in Segment 2.

In Segment 3, changes occur again but in the opposite direction. There is a decrease in the unabsorbed flux of the blackbody, while the temperature increases again. Additionally, the local absorption and the soft photons radius site decrease.

\subsubsection{Emission lines}

\begin{figure*}
\centering
\subfigure{\includegraphics[trim={0cm 0cm 0cm 0cm},width=1\textwidth]{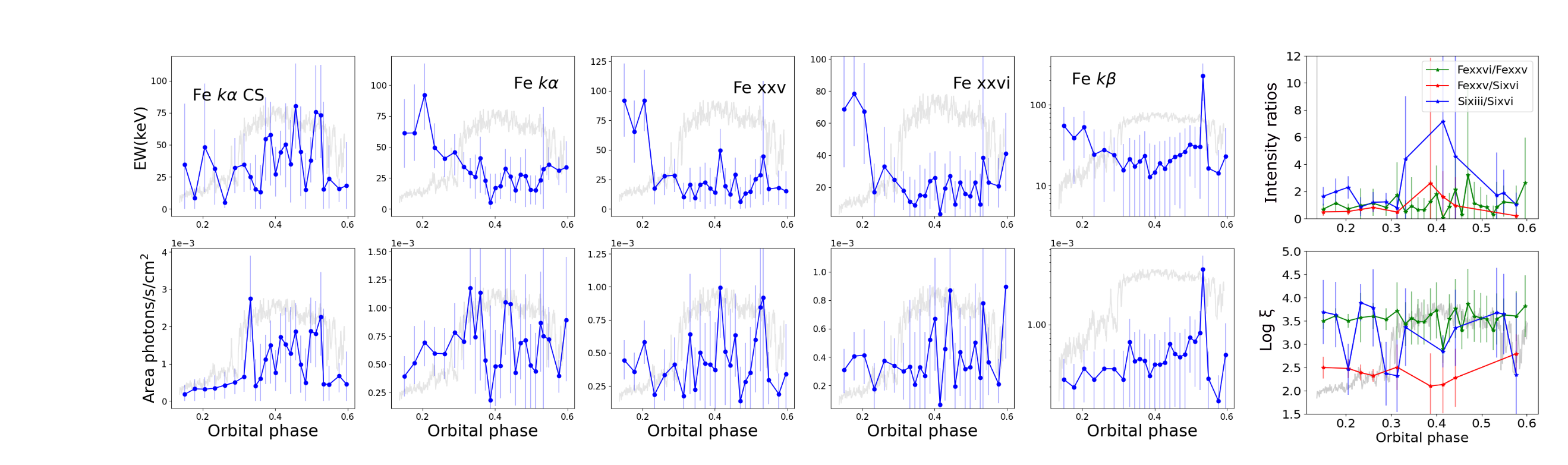}}
\caption{Evolution of the ew (upper panels) and Gaussian fluxes (bottom panels) of emission lines found in the spectra. 
Right panels: line flux ratios and $\log\xi$ parameter, following the analysis performed by \citet{8153132}.
}
\label{feall_lines}
\end{figure*}


A wealth of emission lines from Mg, Si, S, Ca, Cl, Ar and Fe can be observed in the Cen X-3 spectrum (Fig. \ref{fig:lines_example}). We have fitted them all with Gaussians to extract their individual parameters. Fig. \ref{feall_lines}, top panel, represents the evolution of the equivalent width (EW) and intensity of some emission lines. A table with the spectral parameters for the rest of emission lines detected in the spectra for each segment is presented in the Appendix on Table \ref{appendix:B} and Fig. \ref{appendix:C} respectively.

\begin{figure}
\centering
\subfigure{\includegraphics[trim={0cm 7cm 0cm 0cm},width=1\columnwidth]{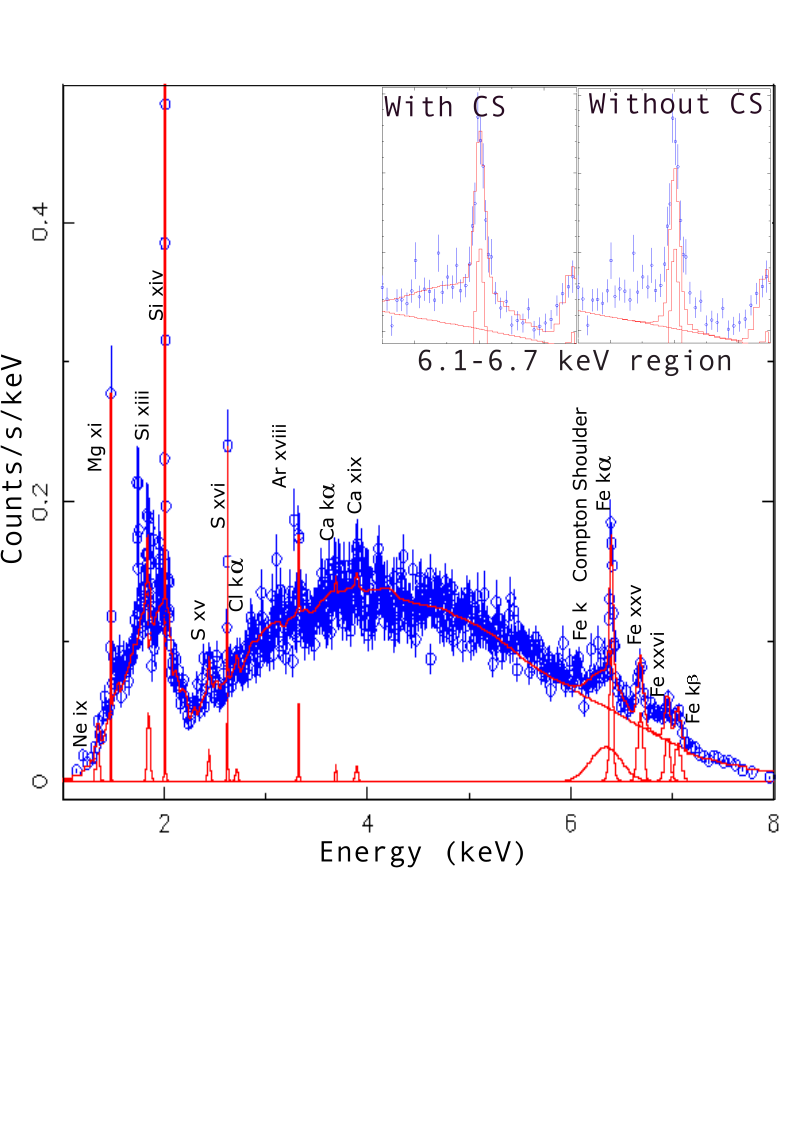}}
\caption{Segment 1-a spectra (blue) plus the model (red) and emission lines detected within this observation. Two inserts in the top right of the Fig. represent the Fe K$\alpha$ region, with and without the Compton shoulder component added to the model.}
\label{fig:lines_example}
\end{figure}

We found a positive correlation between the Fe K$\beta $ and both \ion{Fe}{xxv} and \ion{Si}{xiv} intensity ($\sim$ 0.5). A positive correlation is also observed between the  local absorption column and the Cl K$\alpha $ intensity ($\sim0.64$) and between \ion{Si}{xiv} and the black body temperature ($\sim0.6$). 
The most prominent lines in the \textsc{hetg} spectra of Cen X-3 correspond to Fe (Fig. \ref{feall_lines}, three left-middle and left-bottom panels). The equivalent width (EW, upper panel) of the three lines decrease as the orbital phase increases, a behavior followed by most other emission lines. This is expected, as the intensity of the continuum increases. 

None of the lines show an EW maximum at around $\phi \sim$ 0.25 or the minimum at $\phi \sim$ 0.5, as predicted by the analysis in Paper 1, although this last phase is already observed in a high state. Thus, our hypothesis that large scale dense and cold structures emerging during the egress give rise to these lines, and therefore they should have a maximum and a minimum at orbital phases 0.25 and 0.5 respectively, corresponding to the maximum and minimum projected areas of the flow, cannot be confirmed with the present observation.

Interestingly, the Fe\,K$\alpha$ line, shows a flux enhancement at the low energy wing (see insert on Fig. \ref{all_spectra_spec} and Fig. \ref{fig:lines_example}). This is interpreted as a Compton shoulder. A Compton shoulder is produced through Compton scattering, where high-energy X-ray photons interact with electrons, resulting in a shift of the photon energy. It is more intense, during the phase of high  $N_{\rm H}$, after the change to high state. We have modeled it with an additional Gaussian. The centroid energy and its width $\sigma$ are in the  ranges $[6.33,6.40]$ keV and $[0.05, 0.15]$ keV respectively. These values are compatible with Compton shifts $\Delta E=E^{2}(1-\cos\theta)/mc^{2}+E(1-\cos\theta)$ for a range of scattering angles $\theta$. The maximum energy shift is produced for head on collisions ($\theta=\pi$ rad), for which $\Delta E=0.156$ keV.


Line intensity ratios can be used to measure the ionization state of the emitting plasma. For that purpose we use the ionization parameter: $\log\xi$, calculated by \citet[their Fig. 8, upper panel]{8153132}. We calculated for each interval, separately, the ratio and its corresponding $\log\xi$ (see Fig. \ref{feall_lines} lower right panel), then we computed for each Segment separately its weighted average to the error. The result can be seen on Table \ref{table:ionization_parameter}. 

\begin{table}
\centering
\caption{Ionization parameter ($\log\xi$) for each Segment.}
\begin{adjustbox}{max width=\columnwidth}
\begin{tabular}{ccccc}
\hline\hline
&Segment 1-a &	Segment 1-b &	Segment 2 & Segment 3 \\
\hline
& $\log\xi$ &	 $\log\xi$ &	 $\log\xi$ &	 $\log\xi$ \\
\hline

\ion{Fe}{xxvi}/\ion{Fe}{xxv} & 3.6 $\pm$ 1.2&	3.7 $\pm$ 1.0&	3.6 $\pm$ 0.4&	3.4 $\pm$ 1.5\\

\ion{Fe}{xxvi}/\ion{Si}{xiv} & 2.4 $\pm$ 1.2&	2.5 $\pm$ 1.3&	2.2 $\pm$ 0.7&	2.8 $\pm$ 1.5\\

\ion{Si}{xiii}/\ion{Si}{xiv} & 2.7 $\pm$ 1.0 &	3.3 $\pm$ 0.7&	3.3 $\pm$ 0.6&	3.8 $\pm$ 1.0\\

\hline
\end{tabular}
\end{adjustbox}
\label{table:ionization_parameter}
\tablefoot{Values deduced by the ratios of emission lines found in the spectra following the analysis performed by \citet{8153132}. See Fig. \ref{feall_lines} right panels for their representation.}
\end{table}

In all segments the \ion{Fe}{xxvi}/\ion{Fe}{xxv} ratio indicates the presence of highly ionized plasma.
The \ion{Fe}{xxvi}/\ion{Si}{xiv} ratio point to the presence of a intermediate ionized plasma ($\xi\sim 2.2-2.8$, and the ratio \ion{Si}{xiii}/\ion{Si}{xiv} points to a intermediate ionized plasma in segment 1-a while, the subsequent segments, show a highly ionized plasma (Fig. \ref{feall_lines}, lower right panels and Table \ref{table:ionization_parameter}).

Some He-like triplets can also be found in the \textit{Chandra} spectra of Cen X-3. In order to analyze them we performed a spectral fit for each of the 3 Segments, as the signal-to-noise ratio was not enough to obtain a reliable fit for each interval. The continuum was modeled with an absorbed powerlaw and the lines were fitted with Gaussian functions. Even when the errors are high, we can obtain the parameters $G(T_{\rm e}) = (i + f)/r$ and $R(n_{\rm e}) = f/i$, (where $T_{\rm e}$ represents the plasma electron temperature, $n_{e}$ the electron density, and $i$,$f$ and $r$ the inter-combination, forbidden and recombination emission lines) in order to obtain a rough estimation of plasma temperature and density \citep{2000A&AS..143..495P}.The best results were obtained for \ion{Mg}{xi}, \ion{Si}{xiii}, and Fe\,K$\beta$ for Segments 1,2 and 3 (see Fig. \ref{triplets} and Table \ref{table:triplets}). 

\begin{table}
\centering
\caption{$G$ and $R$ ratios between fluxes of He-like ions.}
\begin{adjustbox}{max width=\columnwidth}
\begin{tabular}{cccc}

\hline\hline
$R$ &Segment 1 &	Segment 2 &	Segment 3 \\
\hline

\ion{Mg}{xi} &	0.1$^{+0.3}_{-0.1}$ &	0.02$^{+0.3}_{-0.1}$ &	0.5$^{+0.6}_{-0.5}$ \\\\
\ion{Si}{xiii} &	$0.7^{+0.3}_{-0.4}$ &	1.1$^{+0.8}_{-0.7}$ &	-\\\\
\ion{Fe}{xxv} &	-&	0.0-0.5&	0.6$\pm 0.5$ \\\\
\hline\hline  			
$G$ &Segment 1 &	Segment 2 &	Segment 3 \\

\hline  

\ion{Mg}{xi}&	$6^{+7}_{-6}$ &-	&	$4^{+6}_{-4}$ \\\\
\ion{Si}{xiii}&	- &	$6^{+7}_{-6}$  &	$2^{+4}_{-2}$ \\\\
\ion{Fe}{xxv} &	$2.3\pm 1.8$ &7-30	&	$2^{+3}_{-2}$ \\

\hline

\end{tabular}
\end{adjustbox}
\tablefoot{Lines and model are represented in Fig. \ref{triplets}. used to estimate plasma temperature and density \citep{2000A&AS..143..495P}.}
\label{table:triplets}
\end{table}

\begin{figure}
\includegraphics[trim={0cm 0cm 0cm 0cm},width=1\columnwidth]{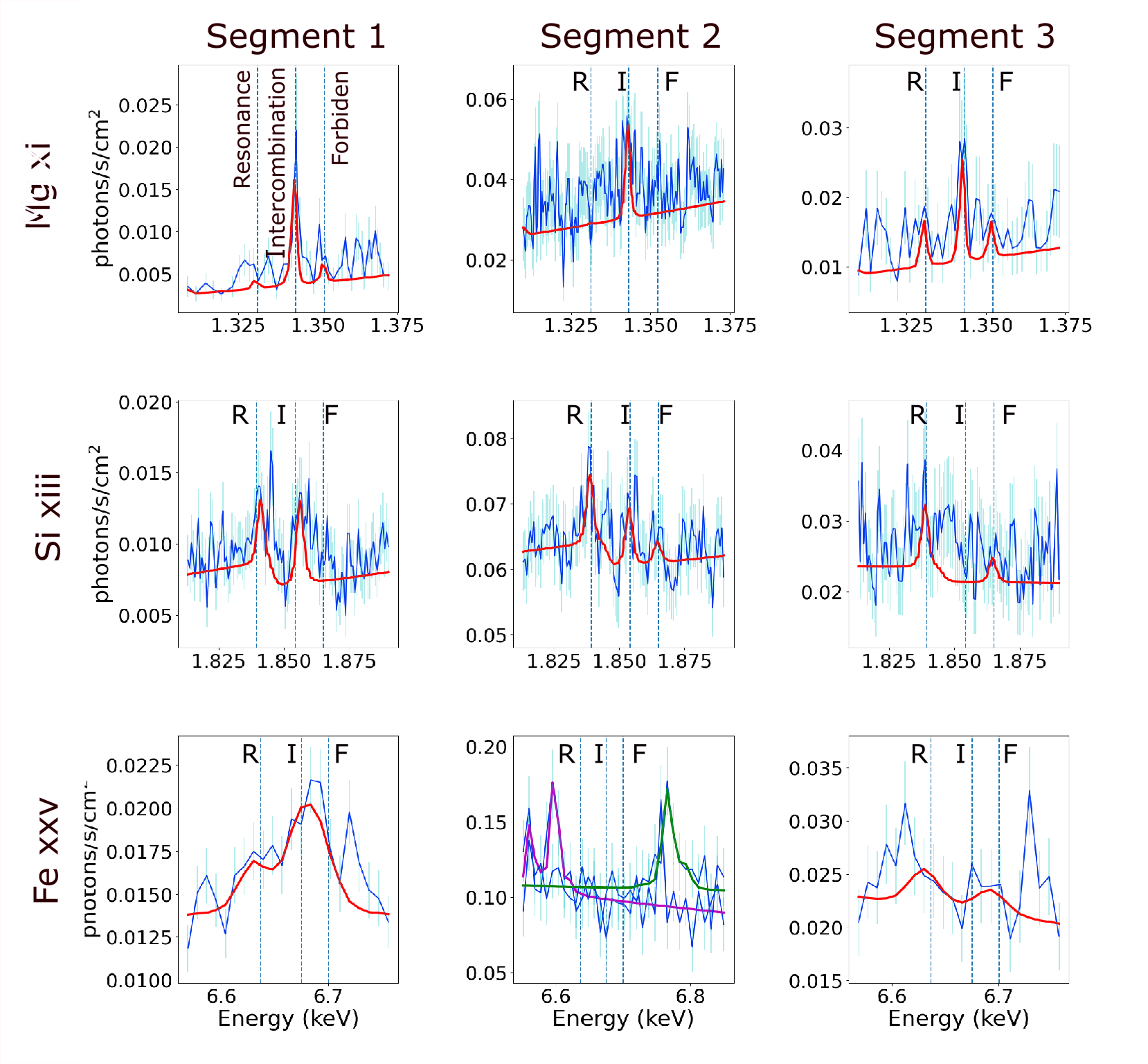}
\caption{He-like ions for each segment (blue) and their models (red). From top to bottom: \ion{Mg}{xi}, \ion{Si}{xiii}, and \ion{Fe}{xxv}. Each column represents a different segment of the observation (Fig. \ref{lcurve}). For \ion{Fe}{xxv} in segment 2, two models were utilized due to the significant deviations from the rest frame energy observed in this emission line. The green model corresponds to blue-shifted spectra, while the magenta model corresponds to red-shifted spectra (as depicted by blue stars and red stars spectral bins, respectively, in Fig. \ref{dopplerfexxv}). The results of the ratio analysis conducted with these fits are summarized on Table\ref{table:triplets}.}
\label{triplets}
\end{figure}

The $G$ parameters of the \ion{Mg}{xi} and \ion{Si}{xiii} ions in Segment 3 suggest temperatures around $0.7\times 10^{(6-7)}$ K. The $R$ parameters of \ion{Mg}{xi} and \ion{Si}{xiii} points to electron densities of $(0.20-1.15)\times 10^{14}$ cm$^{-3}$ in Segment 1, $0.3\times 10^{(13-14)}$ cm$^{-3}$ in Segment 2 and $(0.15-0.50)\times 10^{13}$ cm$^{-3}$ in Segment 3, respectively.

The ratio between Fe K$\beta$ and Fe K$\alpha$ should be $0.13-0.14$ \citep{2003A&A...410..359P}. However, in our observation, the ratio is always significantly higher and close to one in most intervals (0.96 $\pm$ 0.5). This implies that there is a significant number of iron atoms with electrons in both the L and M shells. 

During segment 1 the \ion{Si}{xiii} triplet appears to be blue shifted (Fig. \ref{triplets} left middle panel). We have calculated the expected Doppler shift of the line, assuming that it is moving along the orbit of the compact object. In segment 1 the Doppler shift would be compatible with a motion around the orbit if the emission line production site was on average at $\phi \sim$ 0.17, roughly compatible with the position of the NS in the orbit (provided that the weighted mean of the light curve count rate orbital phase in segment 1 is 0.15), and in segment 2 would be compatible with $\phi \sim 0.50-0.55$ or at rest. In segment 1 the shift is compatible with a velocity component towards the observer of $290-360$ km s$^{-1}$.

More interestingly, the evolution of the central energy of \ion{Fe}{xxv} shows a very clear oscillatory behavior (Fig. \ref{dopplerfexxv}). The \textsc{hetg} spectrum lacks sufficient counts to allow a full high-resolution phase-resolved analysis of the triplet for each interval. We have therefore fitted a single unresolved line. However, after switching to high flux in segment 2, we extracted the average spectrum corresponding to all intervals that showed a blue shift (Fig. \ref{dopplerfexxv} blue stars) on the one hand, and all intervals that showed a red shift (Fig. \ref{dopplerfexxv} red stars) on the other. The resulting spectra (shown in magenta for red shifts and green for blue shifts in the middle bottom panel of Fig. \ref{triplets}) clearly show a shift consistent with the behavior of the unresolved line. This confirms our finding. The small number of counts due to its short duration prevents us from repeating this analysis for segment 3, although the oscillation of \ion{Fe}{xxv} is still present. If interpreted as Doppler shifts, this suggests a coherent motion of the plasma within the system (see section \ref{sec:discussion}).

\begin{figure*}
\includegraphics[trim={0cm 0cm 0cm 0cm},width=1\textwidth]{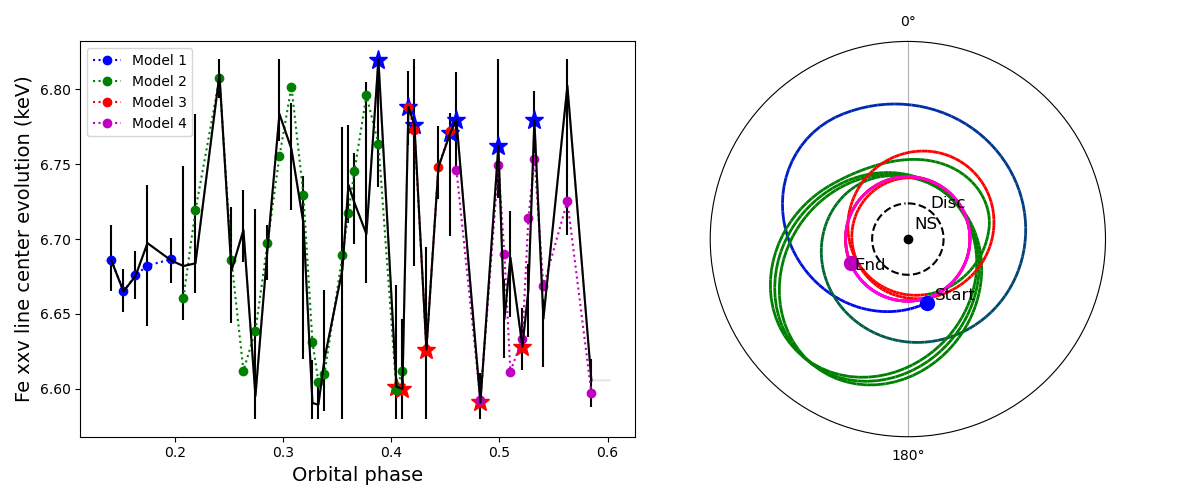}

\caption{\ion{Fe}{xxv} Doppler shifts and trajectory. Left: Observed Doppler shift evolution of the \ion{Fe}{xxv} line center through the observation (black) and the fitted models (blue, green, red, and magenta). The orbital phase refers to the binary system. Right: Projection on the orbital plane of a 3D ballistic trajectory of the \ion{Fe}{xxv} emitting plasma blob around the NS reconstructed from observations. The colors represent the 4 different models as described in Table \ref{table:plasma_orbital_parameters} and the left panel.}

\label{dopplerfexxv}
\end{figure*}

\section{Discussion}
\label{sec:discussion}

The rich phenomenology described above can be understood as an episode of sudden plasma cooling during the transition. For the accretion onto the NS to be effective, the plasma must cool down. This is because ionized matter tangles easily with the magnetic field lines and is expelled out. As the accretion rate increases, the falling plasma temperature, as measured by the blackbody temperature, grows. This is because the falling plasma increases in density and hence the cooling rate. A sudden increase in the mass accretion rate (for example, the arrival of a clump) can trigger a much more effective cooling, resulting in a decrease in the blackbody temperature. This hypothesis is supported by the observation of some high-depth ratio dips ($>1$) in the light curve, which are consistent with clump signatures, both before and during the transition (see Fig. \ref{dip_orbit}). 

The temperature drops from 2.50 keV down to 1.34 keV ($\sim$ by 10 MK) in just two time bins ($\sim 10000$ s). At the same time, the blackbody radius increases from 5 km to 20 km, that is, from a large hot spot to the entire NS, or a boundary layer, as the ingested clump reaches the NS surface. 

We can estimate the amount of matter accreted during the transition (segment 1-b) and afterwards, during segment 2 (Fig. \ref{lcurve}), making use of the change in luminosity. The accretion luminosity can be approximated by the following equation:
\begin{equation}
L_{acc} = \frac{G M_{NS} \dot{M}_{acc}}{R_{NS}}
\end{equation}
A significant change in flux occurs during the transition, in just 10.5 ks, resulting in an approximately threefold increase in luminosity. This corresponds to an accreted mass of $\sim 7.6 \times 10^{20}$ g. Over the entire duration of segment 2, the total estimated accreted mass is approximately $2.7 \times 10^{21}$ g.

During the transition period, six clump candidates were observed. However, during segment 2, this number increased to 46 (see Fig. \ref{dip_orbit}). Consequently, an average mass of approximately $1.3\times 10^{20}$ g and $5.8\times 10^{19}$ g per clump, respectively, is deduced for each segment. These values are comparable to those previously reported \citep[][for Vela X-1]{2010A&A...519A..37F,2014A&A...563A..70M}.

During the observation the local absorption column is remarkably high, ranging from $4\times 10^{23}\text{cm}^{-2}$ during Segment 1, to $2.3\times 10^{24} \text{cm}^{-2}$ during Segment 2. For comparison, the column densities measured during the \textit{XMM-Newton} observations reported on Paper I are at least an order of magnitude lower.

Since the system has a high orbital inclination, the local absorption column varies throughout the orbit. We calculated the theoretical local absorption column depending on the orbital phase. To do so we integrated the stellar wind density, estimated through the Castor, Abbot and Klein (CAK) model assuming an acceleration parameter $\beta=0.8$ \citep{1975ApJ...195..157C} through the path transited by the radiation emitted towards the observer as a function of the orbital phase. This path depends on the major axis, the eccentricity, inclination and argument of the periapsis. The values used to perform this estimation were taken from \cite{falanga}, except for the inclination where we used $74^{\rm o}$, deduced in this work by the evolution of the NS pulse due to Doppler effect generated by its movement through the orbit. The result is that the maximum value for the absorption column through this orbit should be $\sim 28 \times 10^{22}\,\text{cm}^{-2}$ for phases close to the eclipse. Therefore, the observed large absorption column, $\sim 10^{(23-24)}\,\text{cm}^{-2}$, must be produced in the vicinity of the NS. This is also compatible with the high density suggested by the $G$ and $R$ ratios (see Table \ref{table:triplets}). This explains the presence of a Compton shoulder. 

Remarkably, the clump contribution to each interval (see Fig. \ref{den_sec}) exhibits a robust Spearman correlation with the intensity of the Compton shoulder (0.6), and a moderate positive correlation with both the intensity of the Fe K$\beta$ line (0.5) and the local absorption column (0.4), the first panel in Fig  \ref{all_spectra_par} and the first of the bottom panels if Fig. \ref{feall_lines}). This lends support to the hypothesis that the Fe K$\alpha$ lines 
mostly originate in dense and cold segments of the stellar wind (clumps) which concurrently contribute to the local absorption.

It is clear, therefore, that the general configuration of the accreting matter during this observation differs from that during the \textit{XMM-Newton} observations. This is reflected in the different behavior of the Fe lines. 

The central energies of the Fe emission lines show significant departures from their laboratory rest frame values. In particular the \ion{Fe}{xxv} line shows an oscillating central energy. This oscillation is compatible with the movement of the emitting plasma around the NS star in an elliptical trajectory. The parameters of this trajectory change through the observation. We model the resultant Doppler shift through the following equations which also take into account the NS orbital motion around the donor star:

\begin{equation}
r_{d} = \frac{a_{d} \cdot (1 - e_{d}^2)}{1 + e_{d} \cdot \cos(\phi_{d} - W_{d})}; r_{ns} = \frac{a_{ns} \cdot (1 - e_{ns}^2)}{1 + e_{ns} \cdot \cos(\phi_{ns} - W_{ns})}
\label{radius}
\end{equation}

\begin{equation}
v_{\rm D}=-(r_{d}\omega_{d} \sin\phi_{d} \sin i_{d}+r_{ns}\omega_{ns} \sin\phi_{ns} \sin i_{ns})\\
\label{doppler_vel}
\end{equation}

\begin{equation}
\lambda_{\rm D}=\lambda_{rest}\left(1+\frac{v_{\rm D}}{c}\right)\\
\label{doppler_shift}
\end{equation}

where sub-index $d$ and $ns$ refer to the NS orbit around the donor and the emitting plasma trajectory around the NS respectively, $r$ is the orbital radius, $a$ is the semimajor axis, $e$ is the eccentricity, $\phi$ the orbital phase, $W$ the angle to the periapsis, $\omega$ the angular velocity, $i$ the inclination and $\lambda_{\rm D}$ and $\lambda_{rest}$ the center of the emission line, Doppler shifted and at rest, respectively, in wavelength units.

 We utilized only the energy centers of \ion{Fe}{xxv} where their respective areas 
 were incompatible
 with zero. The model was fitted using a \texttt{particle swarm optimization} algorithm (PSO). 
 Our fits revealed that the plasma orbital parameters evolve throughout the observation. Thus, we have divided it into four blocks, corresponding to the four different colors in Fig. \ref{dopplerfexxv}. The fit to a constant value was statistically rejected. On the other hand, a model accounting only for the Doppler shift expected by NS orbital motion, gave significantly poorer results (\textbf{$r^2 \sim 0.005$}). The best fit is achieved through Eqs. \ref{radius}, \ref{doppler_vel}, and \ref{doppler_shift}. The results are shown in Table \ref{table:plasma_orbital_parameters}.

\begin{table}
    \centering
    \caption{Evolution of plasma orbital parameters} 
    \label{table:plasma_orbital_parameters}
    \begin{adjustbox}{max width=\columnwidth}
    \begin{tabular}{lccccc}
        \hline
        \textbf{Model} & \textbf{$r^2$} & \textbf{Semi-Major Axis} &  \textbf{Eccentricity }& \textbf{Period}  & \textbf{Inclination} \\
        &  & \textbf{($R_\star$)} & &\textbf{(s)}  & \textbf{($^{\circ}$)}  \\
        \hline
        
        1 (Blue)	&	0.6	&	0.90 $\pm$ 0.05 &	0.4 $\pm$ 0.3    &	36000 $\pm$ 15000 &	74 $\pm$ 3 \\
        2 (Green)	&	0.7	&	0.91 $\pm$ 0.05 &	0.45 $\pm$ 0.08   &	13000 $\pm$ 130   &	76.0 $\pm$ 0.8 \\
        3 (Red)	    &	1.0	&	0.62 $\pm$ 0.13  &	0.27 $\pm$ 0.14    &	6000 $\pm$ 60     &	81 $\pm$ 6 \\
        4 (Magenta)	&	0.7	&	0.52 $\pm$ 0.02 &	0.003 $\pm$ 0.002&	6300 $\pm$ 50     &	66 $\pm$ 6 \\
    
        \hline
    \end{tabular}
    \end{adjustbox}
    \tablefoot{Values deduced from the model (equations \ref{radius}, \ref{doppler_vel}, and \ref{doppler_shift}) applied to \ion{Fe}{xxv} energy line center Doppler shifts.}
\end{table}

The short time scale on which the transition has occurred (a few tens of ks) is incompatible with a clump or overdense structure traveling through a viscous accretion disk, where the viscosity time is typically comparable to the NS orbital period. We hypothesize that a train of dense stellar wind clumps captured by the NS can initiate the state transition. The emitting plasma must then be moving ballistically. The accretion disk must be within the NS Roche lobe ($\sim 0.3R_\star$). The Keplerian time at the outer radius of the disk is of the order of 70 ks, and any ballistic motion around the NS is expected to be shorter than this time. This behavior is reflected in the oscillations of the \ion{Fe}{xxv} line.

In wind accretion, bound motion of particles around the NS is constrained inside the Bondi-Hoyle radius $R_{B}=2GM/v_{\rm w}^{2}=4\times 10^{10} {\rm cm}/(v_{\rm w}/10^{8} {\rm cm/s})^2$. Even at a moderate relative wind speed of about 500 km s$^{-1}$, the estimated maximum ballistic time around the NS remains in the range of $30-50$ ks, as observed.  While acknowledging the oversimplification inherent in our model, the basic concept of a \ion{Fe}{xxv} emitting plasma in a ballistic rotational trajectory around the NS is the most likely scenario.

\section{Summary and conclusions}
\label{sec:summary}

In this study, we investigated the rich phenomenology observed during the transition period from low-hard to high-soft state in the accreting X-ray binary system Cen X-3. The observed sudden plasma cooling, manifested by a decrease in the blackbody temperature and increase in the density, can be attributed to an effective cooling mechanism allowing high accretion rates. Clump signatures found in the X-ray light curve support this hypothesis, with high-depth ratio dips indicating the presence of clumps triggering more efficient cooling during segment 2. Furthermore, the high local absorption column density and the presence of a Compton shoulder suggest significant differences in accreting matter configuration compared to previous observations.

The oscillations observed in the central energy of the \ion{Fe}{xxv} emission line suggest an elliptical ballistic trajectory of the plasma around the NS. Our analysis, conducted with the high spectral resolution of the \textit{Chandra} observatory, allowed for the modeling of this motion. These findings underscore the promising potential of future missions such as \textit{Athena}, whose higher resolution promises deeper insights into extreme astrophysical phenomena around accreting compact objects.

\section*{Acknowledgements}

This research has been funded by the ASFAE/2022/02 project from the Generalitat Valenciana. N. Schulz and M. Nowak were supported by NASA Chandra grants GO3-24018A and GO3-24018B, respectively. We acknowledge the constructive criticism of the referee whose comments improved the content of the paper.

\section*{Data Availability}
The data analyzed in this study can be found in the {\it Chandra} archive under the observation identification number 26512.



\bibliographystyle{aa}
\bibliography{bib.bib} 



\newpage

\onecolumn

\begin{appendix}

\section{Phase-resolved evolution of the spectral parameters.}

\begin{table*}
\centering
\caption{Phase-resolved spectral parameters for the continuum for each interval.}
\label{Appendix:A}
\begin{adjustbox}{max width=0.8\textwidth}
\begin{tabular}{cccccccccc}
\hline\hline
\\

	Spectra  & Orb. phase 	&	$ \chi ^2 $	&	 $N_{\rm H}^{\rm LOC}$	& $C$ 	&	$N_{\rm H}^{\rm ISM}$	&		$K_{\rm bb}$	 	& $kT_{\rm bb}$	\\\\
 
 	  &	&	&	 ($\times10^{22}$ cm$^{-2}$)	&  	&	($\times10^{22}$ cm$^{-2}$)	&		($\times10^{-3}$) 	 	& (keV))	\\
 \hline

1.1	&	0.14	&	1.2	&	27 $^{+7}_{-6}$    & 1.00 $^{+0.01}_{-0.10}$       & 1.9 $^{+0.0}_{-0.08}$    & 2.7 $\pm$ 0.2   & 1.80 $^{+0.06}_{-0.05}$\\\\
1.2	&	0.18	&	1.2	&	38 $^{+10}_{-8}$   & 1.00 $^{+0.01}_{-0.10}$       & 1.5 $^{+0.2}_{-0.1}$  & 3.40 $^{+0.27}_{-0.24}$  & 1.60 $\pm$ 0.05\\\\
1.3	&	0.21	&	1.2	&	47 $^{+9}_{-7}$    & 1.00 $^{+0.01}_{-0.06}$        & 1.30 $^{+0.10}_{-0.02}$ & 4.10 $^{+0.26}_{-0.24}$  & 1.60 $\pm$ 0.05\\\\
1.4	&	0.23	&	1.5	&	40 $^{+5}_{-4}$    & 1.00 $^{+0.01}_{-0.02}$       & 1.30 $^{+0.03}_{-0.01}$    &7.0 $\pm$ 0.4 & 2.00 $\pm$ 0.05\\\\
1.5	&	0.26	&	1.4	&	50 $^{+6}_{-5}$    & 1.00 $^{+0.01}_{-0.03}$       & 1.30 $^{+0.01}_{-0.01}$    & 11.0 $^{+1.0}_{-0.8}$  &2.4 $\pm$ 0.1\\\\
1.6	&	0.29	&	1.5	&	26$\pm$3           & 1.00 $^{+0.01}_{-0.02}$       & 1.4 $\pm$ 0.1      & 11.0 $\pm$ 0.5 & 2.00 $\pm$ 0.05\\\\
1.7	&	0.31	&	1.1	&	100 $^{+0}_{-13}$  & 1.00 $^{+0.01}_{-0.16}$        & 1.30 $^{+0.04}_{-0.01}$    & 25.0 $\pm$ 0.7  & 1.50 $^{+0.02}_{-0.03}$\\\\
2.1	&	0.33	&	1.3	&	69 $^{+9}_{-8}$    & 1.00 $^{+0.01}_{-0.04}$       & 1.30 $^{+0.01}_{-0.01}$    & 26.0 $\pm$ 0.7  & 1.20 $\pm$ 0.02\\\\
2.2	&	0.35	&	1.2	&	69 $^{+11}_{-9}$   & 1.00 $^{+0.01}_{-0.07}$       & 1.30 $^{+0.03}_{-0.01}$    & 26.0 $\pm$ 0.8  & 1.20 $\pm$ 0.02\\\\
2.3	&	0.36	&	1.1	&	69 $^{+9}_{-8}$    & 1.00 $^{+0.01}_{-0.05}$       & 1.30 $^{+0.01}_{-0.01}$    &27.0 $\pm$ 0.7  & 1.20 $\pm$ 0.02 \\\\
2.4	&	0.37	&	1.2	&	61 $^{+7}_{-6}$    & 1.00 $^{+0.01}_{-0.03}$        & 1.30 $^{+0.01}_{-0.01}$   &26.0 $\pm$ 0.6  & 1.20 $\pm$ 0.02\\\\
2.5	&	0.39	&	1.1	&	66 $^{+8}_{-7}$    & 1.00 $^{+0.01}_{-0.03}$       & 1.30 $^{+0.01}_{-0.01}$   & 27.0 $\pm$ 0.6  & 1.20 $\pm$ 0.02 \\\\
2.6	&	0.4    	&	1.1	&	75 $^{+9}_{-8}$    & 1.00 $^{+0.01}_{-0.03}$       & 1.30 $^{+0.01}_{-0.01}$   & 29.0 $\pm$ 0.67 &1.20 $\pm$ 0.02 \\\\
2.7	&	0.41	&	1.2	&	68 $^{+8}_{-7}$    & 1.00 $^{+0.01}_{-0.03}$       & 1.30 $^{+0.01}_{-0.01}$   & 27.0 $\pm$ 0.6  & 1.20 $\pm$ 0.02\\\\
2.8	&	0.43	&	1.1	&	68 $^{+8}_{-7}$    & 1.00 $^{+0.01}_{-0.03}$       & 1.30 $^{+0.01}_{-0.01}$    &27.0 $\pm$ 0.7  & 1.20 $\pm$ 0.02 \\\\
2.9	&	0.44	&	1.2	&	77 $^{+9}_{-8}$    & 1.00 $^{+0.01}_{-0.03}$        & 1.30 $^{+0.01}_{-0.01}$   &27.0 $\pm$ 0.66 &1.20 $\pm$ 0.02 \\\\
2.1	&	0.46	&	1.2	&	65 $^{+8}_{-7}$    & 1.00 $^{+0.01}_{-0.03}$       & 1.30 $^{+0.01}_{-0.01}$   & 27.0 $\pm$ 0.6  & 1.20 $\pm$ 0.02 \\\\
2.11&	0.47	&	1.1	&	61 $^{+7}_{-6}$    & 1.00 $^{+0.01}_{-0.03}$       & 1.30 $^{+0.01}_{-0.01}$   & 24.0 $\pm$ 0.6  & 1.20 $\pm$ 0.02\\\\
2.12&	0.48	&	1.1	&	63 $^{+9}_{-7}$    & 1.00 $^{+0.01}_{-0.04}$       & 1.30 $^{+0.01}_{-0.01}$   & 24.0 $\pm$ 0.7 & 1.20 $\pm$ 0.02\\\\
2.13&	0.5     &	1.1	&	74 $^{+10}_{-8}$   & 1.00 $^{+0.01}_{-0.04}$       & 1.30 $^{+0.01}_{-0.01}$    & 26.0 $\pm$ 0.62 &1.20 $\pm$ 0.02\\\\
2.14&	0.51	&	1.2	&	71$^{+10}_{-8}$    & 1.00 $^{+0.01}_{-0.03}$       & 1.30 $^{+0.01}_{-0.01}$   & 25.0 $\pm$ 0.7  & 1.20 $\pm$ 0.02\\\\
2.15&	0.53	&	1.1	&	66 $^{+8}_{-7}$    & 1.00 $^{+0.01}_{-0.03}$       & 1.30 $^{+0.01}_{-0.01}$   & 25.0 $\pm$ 0.6  & 1.20 $\pm$ 0.02 \\\\
2.16&	0.54	&	0.8	&	69 $^{+27}_{-18}$  & 1.00 $^{+0.001}_{-0.21}$         & 1.30 $^{+0.06}_{-0.01}$    & 23.0 $^{+1.6}_{-1.5}$  & 1.20 $\pm$ 0.05\\\\
3.1	&	0.55	&	1.1	&	52 $^{+8}_{-6}$    & 0.99 $^{+0.01}_{-0.20}$  & 1.30 $^{+0.04}_{-0.01}$  & 18.0 $^{+1.0}_{-0.7}$  & 1.40 $^{+0.04}_{-0.024}$\\\\
3.2	&	0.58	&	1.2	&	22 $^{+3}_{-2}$    & 1.0 $^{+0.01}_{-0.04}$       & 1.30 $^{+0.08}_{-0.01}$     & 12.0 $^{+0.4}_{-0.3}$  & 1.60 $\pm$ 0.02\\\\
3.3	&	0.6   	&	1.1	&	85 $\pm$ 15        &1.00 $^{+0.01}_{-0.20}$         & 1.30 $^{+0.07}_{-0.01}$   & 23 $\pm$ 1 & 1.30 $^{+0.03}_{-0.04}$\\\\

\hline
\end{tabular}
\end{adjustbox}
\label{appendix:A}
\tablefoot{The first column refers to intervals as shown in Fig. \ref{lcurve}.}
\end{table*}

\section{Emission lines spectral properties for each segment.}

\begin{table*}
 \caption{Parameters of the emission lines model.}
  \centering
  \resizebox{0.33\textwidth}{!}{
  \rotatebox{90}{%

    \begin{tabular}{ccccccccccccccccc}
    
Spectra & \ion{Ne}{ix} &\ion{Mg}{xi} &\ion{Mg}{xii}&\ion{Si}{xiii} & \ion{Si}{xiv}& \ion{S}{xv}& \ion{S}{xvi} &Cl\,K$\alpha$  	&\ion{Ar}{xviii}&	Ca\,K$\alpha$ &\ion{Ca}{xix}&	 Fe\,K$\alpha$ &CS &\ion{Fe}{xxv}&	\ion{Fe}{xxvi}&	 Fe\,K$\beta$\\\\

Center (keV) &0.922	&1.352	&1.473	&1.865	&2.007	&2.461	&2.623&2.69	&3.323	&3.702	&3.902	&6.4	&6.38	&6.7	&6.973&7.058	\\

\hline    
Area&	 & &	 &	 &	 &	 &	 &	 &	 &	& & & & & &\\\\

Segment 1-a &10$\pm$ 3&	0.23$^{+0.23}_{-0.16}$ &0.15$^{+0.15}_{-0.13}$ &0.12$^{+0.03}_{-0.03}$ &0.23$^{+0.02}_{-0.02}$ &0.07$^{+0.03}_{-0.03}$ &0.17$^{+0.04}_{-0.04}$ &0.03$^{+0.03}_{-0.03}$ &0.06$^{+0.02}_{-0.02}$ &0.01$^{+0.02}_{-0.01}$ &0.02$^{+0.03}_{-0.02}$ &0.47$^{+0.06}_{-0.06}$ &1.00$^{+0.13}_{-0.13}$ &0.45$^{+0.06}_{-0.06}$ &0.40$^{+0.07}_{-0.07}$ &0.37$^{+0.07}_{-0.07}$\\\\

Segment 1-b &11$\pm$ 5 &	0.7$^{+0.7}_{-0.5}$ &0.21$^{+0.21}_{-0.15}$ &0.05$\pm$ 0.04&	0.27$\pm$ 0.05&-&	0.17$^{+0.03}_{-0.1}$ &0.27$^{+0.13}_{-0.13}$ &0.14$\pm$ 0.1&	0.00$^{+0.02}_{-0.01}$ &0.06$^{+0.08}_{-0.06}$ &0.59$^{+0.12}_{-0.12}$ &1.8$\pm$ 0.3&	0.45$\pm$ 0.13&	0.35$\pm$ 0.13&	0.57$\pm$ 0.15\\\\
									
Segment 2 &13$\pm$ 3&	1.1$^{+1.1}_{-1.0}$ &0.8$^{+0.8}_{-0.6}$ &0.4$\pm$ 0.1&	- &	0.4$\pm$ 0.6&	0.14$\pm$ 0.12&	0.1$\pm$ 0.1&	- &	- &	- &	0.5$\pm$ 0.1&	3.46$\pm$ 0.21 &	0.5$\pm$ 0.8&	0.56$\pm$ 0.08&	0.81$\pm$ 0.09\\\\

Segment 3 &47 $\pm$ 13 &	0.6$^{+0.6}_{-0.4}$ &0.19$^{+0.19}_{-0.13}$ &0.13$\pm$ 0.07&	0.14$\pm$ 0.05 & 0.19$\pm$ 0.10 &	-&	0.1$\pm$ 0.7&	0.06$\pm$0.06&	0.16$\pm$0.10&	0.05$^{+0.06}_{-0.05}$ &0.64$\pm$ 0.13 &	0.97$\pm$  0.03 &	0.15$\pm$ 0.12 &	0.20$\pm$ 0.13 &	0.27$\pm$ 0.13\\\\
									
\hline
eqw & & & & & & & & & & & & & & & & \\
\hline
Segment 1-a & 2100 $\pm$ 600 & 40$^{+4}_{-12}$ &  23 $\pm$ 4 & 17 $\pm$ 4 & 33 $\pm$ 3 & 10 $\pm$ 5 & 26 $\pm$ 5 & 4 $\pm$ 4 & 8 $\pm$ 3 & 1.3$^{+2.3}_{-1.3}$ & 3 $\pm$ 3 & 58 $\pm$ 8 & 177 $\pm$ 23 & 82 $\pm$ 12 & 80 $\pm$ 14 & 76 $\pm$ 15 \\\\

Segment 1-b & 500 $\pm$ 220 & 27 $\pm$ 8 & 5.5 $\pm$ 1.6 & 1.6 $\pm$ 1.3 & 9.1 $\pm$ 1.6 & - & 7.8 $\pm$ 3.1 & 9 $\pm$ 4 & 5 $\pm$ 3 & - & 2$^{+3}_{-2}$ & 32 $\pm$ 7 & 135 $\pm$ 22 & 36 $\pm$ 10 & 32 $\pm$ 12 & 54 $\pm$ 14 \\\\

Segment 2 & 170 $\pm$ 40 & 12.5 $\pm$ 2.1 & 1.6 $\pm$ 0.3 & 4.1 $\pm$ 0.9 & - & 3.8 $\pm$ 1.3 & 8.1 $\pm$ 3.2 & 1.2 $\pm$ 1.2 & - & - & - & 16 $\pm$ 3 & 160 $\pm$ 9 & 28 $\pm$ 4 & 36 $\pm$ 5 & 55 $\pm$ 6 \\\\

Segment 3 & 1600 $\pm$ 500 & 17 $\pm$ 6 & - & 3.1 $\pm$ 1.7 & 3.5 $\pm$ 1.1 & 4.6 $\pm$ 2.4 & - & 2.9 $\pm$ 1.7 & 1.5 $\pm$ 1.5 & 4 $\pm$ 3 & 1.5$^{+1.6}_{-1.5}$ & 34 $\pm$ 7 & 53 $\pm$ 17 & 10 $\pm$ 8 & 15 $\pm$ 10 & 21 $\pm$ 10 \\\\
\hline

      \hline
    \end{tabular}

    }
    }
  \label{appendix:B}
\end{table*}

\begin{figure*}
\centering
\subfigure{\includegraphics[trim={1cm 1cm 0cm 0cm},width=1\textwidth]{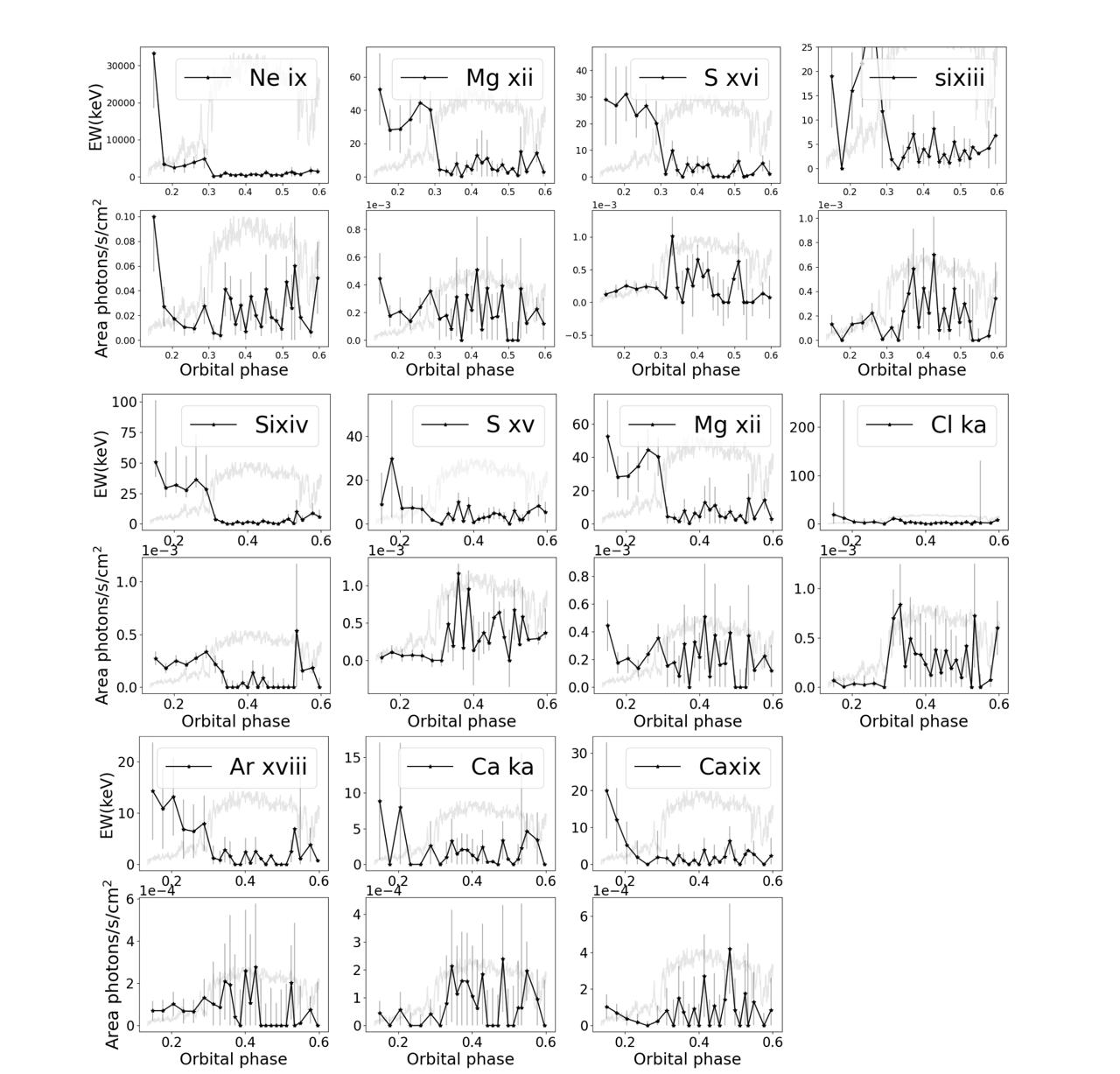}}
\caption{Ew (upper panels) and area (lower panels) of some of the emission lines detected in the spectra.
}
\label{appendix:C}
\end{figure*}

\label{lastpage}
\end{appendix}
\end{document}